\documentclass[12pt]{article}
\usepackage[dvips]{color,graphicx}
\usepackage{epsfig,amsmath}
\usepackage[english]{babel}
\usepackage{pgf,pgfarrows,pgfnodes,pgfautomata,pgfheaps}
\usepackage{amsfonts,amsmath,amsthm,amssymb,enumerate,array,amssymb}
\usepackage[latin1]{inputenc}
\usepackage{bm}
\usepackage{fixmath}
\usepackage{amsbsy}
\usepackage{graphicx}
\usepackage{capt-of}
\usepackage{dcolumn}
\usepackage{hyperref}

\newcommand{\be}{\begin{equation}}
\newcommand{\ee}{\end{equation}}
\newcommand{\bea}{\begin{eqnarray}}
\newcommand{\eea}{\end{eqnarray}}

\begin{document}

\begin{titlepage}

\begin{flushright}
UTTG-09-11
\end{flushright}

\begin{center}
{\Large\textbf{\mathversion{bold}
Kinetic Theory of Collisionless Self-Gravitating Gases:\\ Post-Newtonian Polytropes}\par}
\end{center}

% \vskip 0.1cm
\begin{center}
C. A. Agón$^{\dagger}$\footnote{cesar.agon@nucleares.unam.mx}, J. F. Pedraza$^{\star}$\footnote{jpedraza@physics.utexas.edu} and
J. Ramos-Caro$^{\natural}$\footnote{javier@ime.unicamp.br}

\vspace{0.2cm}
${}^{\dagger}$  Instituto de Ciencias Nucleares, Universidad Nacional
   Aut\'onoma de M\'exico,\\ Apartado Postal 70-543, M\'exico D.F. 04510, M\'exico\\
 \vspace{0.2cm}
 
$^{\star}$ Theory Group, Department of Physics, University of Texas,\\
 1 University Station C1608, Austin, TX 78712, USA\\

 \vspace{0.2cm}
$^{\natural}$ Departamento de Matemática Aplicada, IMECC, Universidade Estadual de Campinas,\\
            Campinas, São Paulo 13083-859, Brazil
\vspace{0.2cm}

\end{center}

{\bf Abstract:} In this paper we study the kinetic theory of many-particle astrophysical systems and
we present a consistent version of the collisionless Boltzmann equation in the 1PN approximation.
 We argue that the equation presented by Rezania and Sobouti in A\&A 354 1110 (2000) is not the correct expression to describe the evolution
 of a collisionless self-gravitating gas. One of the reasons that account for the previous statement is that
the energy of a free-falling test particle, obeying the 1PN equations of motion
for static gravitational fields, is not a static solution of the mentioned equation.
The same statement holds for the angular momentum, in the case of spherical systems.
We provide the necessary corrections and obtain an equation that is consistent with the corresponding
equations of motion and the 1PN conserved quantities. We suggest some potential relevance for the study of high density astrophysical systems and as an application we construct the corrected version of the post-Newtonian polytropes.

\vspace{0.2in}

\smallskip
\end{titlepage}

\setcounter{footnote}{0}

\section{Introduction}

The evolution of self-gravitating systems is a subject of great interest in astrophysics.
From the statistical standpoint, the most straightforward way to construct self-consistent stellar systems
is by means of finding the distribution function (DF) for a stellar system with a known gravitational potential
 and matter distribution. Since the mass density
is the integration of the distribution function over the velocity variable in the phase space of the system, the
problem of finding a DF is that of solving an integral equation (see \cite{eddi,frick,lynden,hunt93,PRG}
 and the references therein). This construction is also the so-called ``from $\rho$ to $f$'' approach for finding a
self-consistent distribution function $f$ \cite{BT}, although the opposite procedure is also used sometimes.

Now, in the framework of the general theory of relativity it is assumed that the DF
satisfies the general relativistic version of the Fokker-Planck equation \cite{kand1,kand2,kand3}
or the collisionless Boltzmann equation (CBE) \cite{kremer,chacon}. The former is devoted to systems
in which local gravitational encounters play an essential role in their evolution whereas the latter is
useful to study systems sufficiently smooth, so that they may be considered to be
collisionless \cite{BT}. One can actually consider systems in which a number of particle species can
collide and produce different species. This is how the formation of the light elements in the big bang
 nucleosynthesis is calculated (see \cite{nucleus} for a review).

However, if we want to study the dynamics of huge astrophysical ensembles such as galaxies and galaxy clusters,
physical collisions between the stars are very rare, and the effect of gravitational collisions can be neglected
for times far longer than the age of the universe.
Those systems are characterized by a relaxation time, $t_{{\rm relax}}$, that
is arbitrarily large in comparison with their crossing time, $t_{{\rm cross}}$, and this means that
they can be approximated as a continuum rather than concentrated into nearly point-like stars. The same
holds (with some restrictions) in the case of collisional
 systems such as globular clusters, neutron stars and withe dwarfs, where the relativistic
 effects of gravitation become important. Although $t_{{\rm relax}}$ here is significantly smaller than the system's age,
the CBE is still valid over periods of time
shorter than $t_{{\rm relax}}$ or when it is recognized
that the system evolves slowly towards the equilibrium
(on a timescale of the order of $t_{{\rm relax}}$).
For example, Taruya \& Sakagami showed in \cite{taruya,taruya2},
that the evolution of spherically symmetric systems in the collisional regime
can be modeled as a sequence of polytropic states (i.e. described by a DF proportional
to $E^{\gamma}$, which is a static solution of the CBE),
with increasing polytropic index.

There are many systems in astrophysics where Newtonian gravity is dominant, but general relativistic gravity plays also
an important role in their evolution. For such systems it would be nice to have an approximation
scheme which gives a Newtonian description in the lowest order and general relativistic effects
as higher order perturbations. The post-Newtonian approximation is perfectly suited for this
purpose. The appropriate scheme that describes the effects of the first post-Newtonian
corrections on the motion of test particles, was first formulated by Einstein and
Infeld \cite{eins1, eins2, eins3}, then revisited by Weinberg \cite{WB} and it is
known as the first post-Newtonian (1PN) approximation. This approach gives the
corrections up to order $\bar{v}^{2}/c^{2}$, where $\bar{v}$ is a typical velocity in the system and $c$ is
the speed of light, and it holds for particles moving non
relativistically ($\bar{v}\ll c$), as in the case of a star moving around a typical
galaxy. Currently, PN approximations to higher orders have been developed
because of the increasing interest around kinematics and associated emission of
gravitational waves by binary pulsars, neutron stars and black holes. The most
promising candidates for detecting such profiles are LIGO, VIRGO and GEO600 (see
\cite{futamase, bichak} for references).

The first attempt to derive a kinetic equation in the 1PN approximation was done a few years ago by Rezania and Sobouti \cite{res}, finding some
relevant solutions. They tried to obtain the post-Newtonian version of the Liouville's
equation for a gas of identical collisionless particles and, as an application, they constructed the 1PN
version of the classical polytropes \cite{eddi}. Strictly speaking, this equation should be called the
collisionless Boltzmann equation since the one attributed to Liouville refers
to an $N$-particle equation. However, both equations can be shown to coincide
in the case $N\gg1$ and when the $N$-body distribution function is separable
(see for example \cite{BT}).

The authors started
from the general relativistic kinetic equation
\begin{equation}\label{liouville}
{\cal L}_U f = \left(U^\mu \frac{\partial}{\partial x^\mu} - \Gamma_{\mu\nu}^i U^\mu
U^\nu \frac{\partial}{\partial U^i}\right) f(x^\mu,U^i) = 0,
\end{equation}
where $(x^\mu,U^i)$ is the set of configuration and four-velocity coordinates, $\Gamma^i
_{\mu\nu}$ are the Christoffel's symbols,
${\cal L}_U$ is the Liouville's operator and  $f(x^\mu,U^i)$ is the
 one particle DF. Then, using the fact that
the four-velocity of the particle satisfies the relation
\begin{equation}\label{U-condition}
    g_{\mu\nu}U^\mu U^\nu=-c^2,
\end{equation}
they performed an expansion of ${\cal L}_U$
up to the order $(\bar{v}/c)^{2}$, where $\bar{v}$ is the typical Newtonian
speed. The resulting post-Newtonian approximation of (\ref{liouville}) was:
\begin{equation}
\begin{split}
&\quad\quad\quad\quad\quad\quad\quad\quad\quad\quad\quad\quad\quad
\frac{\partial f}{\partial t} +v^{i}\frac{\partial f}{\partial x^{i}}
  -\frac{\partial \phi}{\partial x^{i}}\frac{\partial f}{\partial v^{i}}\\
&-\frac{1}{c^{2}}\left[(4\phi+\mathbf{v}^{2})\frac{\partial \phi}{\partial x^{i}}
  -v^{i}v^{j}\frac{\partial \phi}{\partial x^{j}}-v^{i}\frac{\partial \phi}{\partial t}+\frac{\partial \psi}{\partial x^{i}}
  +\left(\frac{\partial \xi_{i}}{\partial x^{j}}-\frac{\partial \xi_{j}}{\partial x^{i}}\right)v^{j}
  +\frac{\partial \xi_{i}}{\partial t}\right]\frac{\partial f}{\partial v^{i}}
  =0,\label{liouville-res}
\end{split}
\end{equation}
where $\phi$ is the Newtonian gravitational potential,
$\psi$ and $\xi_{i}$ are the post-Newtonian potentials and
$f$ is now interpreted as a DF depending on the
spatial coordinates $\mathbf{x}$, the Newtonian velocity $\mathbf{v}$,
and the time $t$.

As it was pointed out by the authors, one can verify that the above equation admits
the static solution ($\partial\phi/\partial t=\partial\psi/\partial t=0$ and $\xi_{i}=0$)
\begin{equation}\label{E-rez}
    E^{*}=\frac{1}{2}\mathbf{v}^{2}+\phi+(2\phi^{2}+\psi)/c^{2},
\end{equation}
which could be interpreted as the 1PN generalization of the classical energy.
Moreover, for the case of spherically symmetric systems, there appear three additional
integrals of equation (\ref{liouville-res}):
\begin{equation}\label{l-rez}
    l_{i}^{*}=\varepsilon_{ijk}x^{j}v^{k}(1-\phi/c^{2}),
\end{equation}
which could also be considered as post-Newtonian extensions of the classical angular momenta.
Thus, DFs depending on the integrals (\ref{E-rez}) and (\ref{l-rez})
would represent the 1PN statistical description of spherically symmetric systems in equilibrium.

However, we can verify that the ``energy'' given by (\ref{E-rez}) is not an integral
of the 1PN equations of motion for a static source of gravitation \cite{WB}
\begin{equation}
\frac{d\mathbf{v}}{dt} = - \nabla \left(\phi + \frac{2 \phi^{2} +
\psi}{c^{2}}\right) + \frac{4 \mathbf{v}}{c^{2}} (\mathbf{v} \cdot \nabla \phi)
- \frac{\mathbf{v}^{2}}{c^{2}} \nabla \phi. \label{1PN-eq-motion}
\end{equation}
%where $\nabla$ is the nabla operator.
Moreover, if we assume spherical symmetry for
the fields $\phi$ and $\psi$, we can also verify that the ``angular momentum''
(\ref{l-rez}) is not an integral of motion of the above equation.
These statements
imply that there is not any correspondence between the microscopic
description of motion given by (\ref{1PN-eq-motion}) and the
macroscopic (statistical) description involved in (\ref{liouville-res}).
Since the former has been demonstrated to be correct
(see \cite{eins1,eins2,eins3,WB,futamase,bichak,chan}  for
references), we can establish that
the equation (\ref{liouville-res}) does not provide an appropriate statistical description for the
evolution of a collisionless gas of self-gravitating particles.
Then, it is necessary to derive a corrected version of this relation.
Such version must admit as a static solution, the correct expression
for the energy at 1PN order (see section \ref{sec:integrals}):
\begin{equation}
\label{energia-main}
E=\frac{\mathbf{v}^{2}}{2}+\phi+\frac{3\mathbf{v}^{4}}{8c^2}-\frac{3\mathbf{v}^{2}\phi}{2c^2}+\frac{\phi^2}{2c^2}+\frac{\psi}{c^2},
\end{equation}
and, in the spherically symmetric case, the 1PN generalization of the angular momentum:
\begin{equation}\label{momento-angular-main}
    l_{i}=\varepsilon_{ijk}x^{j}v^{k}\left(1-\frac{3\phi}{c^{2}}+\frac{\mathbf{v}^{2}}{c^{2}}\right).
\end{equation}

In section \ref{sec:ec-poporra} we show a detailed derivation of the corrected version of the CBE in the 1PN approximation. The final result reads
\be
\begin{split}
&\quad\quad\quad\quad\quad\quad\quad\frac{\partial f}{\partial t}+v^{i}\frac{\partial f}{\partial x^{i}}-
\frac{\partial \phi}{\partial x^{i}}\frac{\partial f}{\partial v^{i}}+
\frac{1}{c^{2}}\left(\frac{\mathbf{v}^{2}}{2}-\phi \right)\left(\frac{\partial f}{\partial t}
+v^{i}\frac{\partial f}{\partial x^{i}}\right)\\
&+\frac{1}{c^{2}}\left[4v^{i} v^{j}\frac{\partial \phi}{\partial x^{j}}-
\left(\frac{3\mathbf{v}^{2}}{2}+3\phi\right)\frac{\partial \phi}{\partial x^{i}}+
3v^{i}\frac{\partial \phi}{\partial t}-\frac{\partial \psi}{\partial x^{i}}-\frac{\partial \xi_{i}}{\partial t}
-v^{j}\left(\frac{\partial \xi_{i}}{\partial x^{j}}-\frac{\partial \xi_{j}}{\partial
  x^{i}}\right)\right]\frac{\partial f}{\partial v^i}=0.\label{liouv-final-main}
\end{split}
\ee
 We can verify that the energy,
 given by (\ref{energia-main}), is a static solution of (\ref{liouv-final-main}), as well as,
  the 1PN angular momentum (\ref{momento-angular-main}), for the spherically
 symmetric case. It is worth to
point out that the corrections introduced above, should be
 taken into account in the analysis of \cite{res2} and \cite{ramos}, since they used
 some of the results of \cite{res}.

The rest of the paper is organized as follows. In Section \ref{sec:integrals} we start by deriving the post-Newtonian integrals of
motion corresponding to static and spherically symmetric gravitational field. Then, in Section
\ref{sec:ec-poporra}, we show a detailed obtention of (\ref{liouv-final-main}), and we rewrite it
in a number of ways (\ref{form1})-(\ref{liouv-alt2}) in order to shed some light on its physical meaning.
Finally, in Section \ref{sec:staticsol}, we construct the corrected version of the post-Newtonian polytropes.
To do so, we employ the ``$f$ to $\rho$'' approach. We start with a DF that is just a generalization of the
classical one and then we integrate over the phase space in order to obtain the different components of the
stress tensor. Then we use the field equations to recover the Newtonian and post-Newtonian potentials,
and we use them to study the main properties of the models.

 \section{Integrals of Motion in the 1PN aproximation}\label{sec:integrals}
We start by showing a detailed derivation of the first integral of motion, i.e. the energy,
for a free-falling test particle in the presence of  static gravitational fields
($\partial\phi/\partial t=\partial\psi/\partial t=0$ and $\xi_{i}=0$),
in the 1PN approximation. In order to do this
we consider the Lagrangian corresponding to the free-falling motion,
\begin{equation}\label{lag0}
2\mathfrak{L}=\;g_{\mu\nu}U^\mu U^\nu,
\end{equation}
where $g_{\mu\nu}$ is the metric tensor. We shall adopt the following conventions:
\begin{equation}\label{convention1}
x^\mu=(ct,x^i),\quad\quad U^\mu=\frac{\partial x^\mu}{\partial\tau}=\frac{\partial
 x^0}{\partial\tau}\frac{\partial x^\mu}{\partial x^0}=U^0V^\mu,
\end{equation}
and
\begin{equation}\label{convention2}
\frac{\partial x^0}{\partial\tau}=U^0,\quad\quad\frac{\partial x^\mu}{\partial x^0}=V^\mu=(1,v^i/c),
\end{equation}
where $\tau$ denotes the particle's proper time, Greek indices range from $0$ to $3$ and Latin indices range from $1$ to $3$.

At 1PN order, the line element $ds^{2}=g_{\mu\nu}dx^{\mu}dx^{\nu}$, can be written
in terms of the Newtonian potential $\phi$ and post-Newtonian potentials $\psi$ and $\xi_i$ defined as in \cite{WB}
\begin{equation}
ds^2=-\left(1+\frac{2\phi}{c^2}+\frac{2(\phi^2+\psi)}{c^4}\right)c^2dt^2+2\left(\frac{\xi_i}{c^3}\right)cdtdx^i+
\left(1-\frac{2\phi}{c^2}\right)\delta_{ij}dx^idx^j.
\end{equation}
Then, for the case of static configurations,
the post-Newtonian Lagrangian (\ref{lag0}) reduces to
\begin{equation}
\label{lag}
2\mathfrak{L}=
-\left(1+\frac{2\phi}{c^2}+\frac{2(\phi^2+\psi)}{c^4}\right)c^2\dot{t}^2
+\left(1-\frac{2\phi}{c^2}\right)\dot{x}^{i}\dot{x}^{j}\delta_{ij},
\end{equation}
where the upper dot denotes derivation with respect to $\tau$.
It is possible to verify that, upon the application of the variational principle,
the corresponding Euler-Lagrange equations lead to the
1PN equations of motion showed in (\ref{1PN-eq-motion}).
Since this Lagrangian is time-independent, the energy
\begin{equation}
E=-\frac{\partial\mathfrak{L}}{\partial\dot{t}}
\end{equation}
is the first integral of motion. Our goal is to compute this quantity up to order $c^{-2}$.
By taking the derivative of the Lagrangian with respect to $\dot{t}$ we obtain then
\begin{equation}\label{E1}
E=\left(1+\frac{2\phi}{c^2}+\frac{2(\phi^2+\psi)}{c^4}\right)c^2\dot{t}.
\end{equation}
Now we have to write $\dot{t}$ in terms of the spatial coordinates.
In order to do this, first remember that $\dot{t}=U^{0}/c$. The quantity $U^{0}$
can be computed by means of equation (\ref{U-condition}) in 1PN approximation,
\begin{equation}
-\left(1+\frac{2\phi}{c^2}+\frac{2(\phi^2+\psi)}{c^4}\right)(U^0)^2
+\left(1-\frac{2\phi}{c^2}\right)\frac{\mathbf{v}^{2}}{c^2}(U^0)^2=-c^2,
\end{equation}
and after some calculations we can find, up to order $c^{-4}$,
\begin{equation}\label{Ucero}
\frac{U^0}{c}=1+\frac{\mathbf{v}^{2}}{2c^2}-\frac{\phi}{c^2}+\frac{3\mathbf{v}^{4}}{8c^4}-
\frac{5\mathbf{v}^{2}\phi}{2c^4}+\frac{\phi^2}{2c^4}-\frac{\psi}{c^4}=\dot{t}.
\end{equation}
Finally, by introducing the above relation in (\ref{E1}) we obtain the
energy at order $c^{-2}$:
\begin{equation}
\label{energia}
E=\frac{\mathbf{v}^{2}}{2}+\phi+\frac{3\mathbf{v}^{4}}{8c^2}-\frac{3\mathbf{v}^{2}\phi}{2c^2}+\frac{\phi^2}{2c^2}+\frac{\psi}{c^2}.
\end{equation}
Note that we have suppressed the constant term $c^2$,
 which is meaningless for our purposes. By taking the total derivative of (\ref{energia})
  with respect to $t$, and using the 1PN equations of motion (\ref{1PN-eq-motion}),
   we can indeed prove that $E$ is a conserved quantity at order $c^{-2}$.
   It turns out that this quantity is also a static solution of
   the corrected CBE (\ref{liouv-final-main}),
    which can be proved by simple inspection (see section
\ref{sec:staticsol}, eq. (\ref{liouville-static})).

Using the 1PN equations of motion we can also rewrite the Lagrangian (\ref{lag}), up to a total derivative, as
\begin{equation}\label{lag-wein}
\mathfrak{\tilde{L}}=\frac{\mathbf{v}^{2}}{2}-\phi-
\frac{\phi^2}{2c^2}-\frac{3\phi \mathbf{v}^{2}}{2c^2}+\frac{\mathbf{v}^{4}}{8c^2}-\frac{\psi}{c^2},
\end{equation}
which in fact agrees with the Lagrangian presented in \cite{WB}. Now, if we assume spherical symmetry,
the fields $\phi$ and $\psi$ will depend on the spatial coordinates only through
$r=\sqrt{(x^{1})^{2}+(x^{2})^{2}+(x^{3})^{2}}$ and, in this case,
the Lagrangian has two cyclic coordinates $\theta$ and $\varphi$
given by
\bea
\tan\varphi&=&x^{2}/x^{1},\\
\cos\theta&=&x^{3}\left[(x^{1})^{2}+(x^{2})^{2}+(x^{3})^{2}\right]^{-1/2}.
\eea
This fact enable us to find that the three quantities
\begin{equation}\label{momento-angular}
    l_{i}=\varepsilon_{ijk}x^{j}v^{k}\left(1-\frac{3\phi}{c^{2}}+\frac{\mathbf{v}^{2}}{2c^{2}}\right),
\end{equation}
are also integrals of motion and to interpret them as the
post-Newtonian generalization of the angular momenta. For axial symmetry,
of course, it is straightforward to show that only the $z$-component of the angular momentum is conserved.

The quantities (\ref{energia}) and (\ref{momento-angular}) are not static
solutions of the equation (\ref{liouville-res}), obtained in \cite{res}, but they will be
integrals of the corrected 1PN CBE, as expected.
The derivation of such equation will be the aim of the next section.

 \section{Derivation of the CBE in the 1PN Approximation}\label{sec:ec-poporra}
In order to obtain a post-Newtonian approximation of equation (\ref{liouville}), it
 is convenient to change from $(x^\mu,U^i)$ to $(x^\mu,v^i)$.
In other words, we use a transformation of the form
\bea
(x^{\mu},U^{i})\rightarrow(x^{\mu},v^{i}(x^{\mu},U^{i})),
\eea
so that the distribution function becomes
\bea
f(x^{\mu},U^{i})=f(x^{\mu},v^{i}(x^{\mu},U^{i})).
\eea
The explicit dependence of $v^{i}=v^{i}(x^{\mu},U^{i})$ can be inferred from
(\ref{U-condition}) and the fact that
$U^i = U^0  v^i/c$ (see (\ref{convention1}) and (\ref{convention2})). This allows us to write
\bea
v^i(x^{\mu},U^{i}) =\frac{cU^i}{U^{0}(x^{\mu},U^{i})}.
\eea
To simplify the calculation
we will first use the variable $V^{\mu}=(1,v^{i}/c)$ and then we will come back to $v^{i}$.
The partial derivatives of $f$ transform as
\bea
\left( \frac{\partial f}{\partial x^\mu}\right)_{U^{i}}&=&
\left( \frac{\partial f}{\partial x^\mu}\right)_{V^{i}}+
\left( \frac{\partial f}{\partial V^j}\right)_{x^{\mu}}
\left(\frac{\partial V^j}{\partial x^{\mu}}\right)_{U^i}\nonumber \\
\left( \frac{\partial f}{\partial U^i}\right)_{x^{\mu}}&=&
\left( \frac{\partial f}{\partial V^j}\right)_{x^{\mu}}\left(\frac{\partial V^j}{\partial U^i}\right)_{x^{\mu}},
\eea
where the subscript at the bottom of the various derivatives indicates the quantity that is constant
when $f$ is differentiated. According to the above relations, equation (\ref{liouville}) is rewritten as
\begin{equation}\label{liouville2}
{\cal L}_U f=U^0 V^\mu\left[\left( \frac{\partial f}{\partial x^\mu}\right)_{V^{i}}+
\left( \frac{\partial f}{\partial V^j}\right)_{x^{\mu}}
\left(\frac{\partial V^j}{\partial x^{\mu}}\right)_{U^i}\right]-\Gamma_{\mu\nu}^i{U^{0}}^{2} V^\mu
V^\nu\left[\left( \frac{\partial f}{\partial V^j}\right)_{x^{\mu}}
\left(\frac{\partial V^j}{\partial U^i}\right)_{x^{\mu}}\right]=0,
\end{equation}
where
\bea
\left(\frac{\partial V^j}{\partial x^{\mu}}\right)_{U^i}&=&-
\frac{U^{j}}{{U^{0}}^{2}}\left(\frac{\partial U^0}{\partial x^{\mu}}\right)_{U^i}, \\
\left(\frac{\partial V^j}{\partial U^i}\right)_{x^{\mu}}&=&\frac{\delta^{ij}}{U^{0}}-
\frac{U^{j}}{{U^{0}}^{2}}\left(\frac{\partial U^0}{\partial U^{i}}\right)_{x^{\mu}}.
\eea
The terms $\left({\partial U^0}/{\partial x^{\mu}}\right)_{U^i}$
and $\left({\partial U^0}/{\partial U^{i}}\right)_{x^{\mu}}$
can be obtained by differentiating equation (\ref{U-condition}). The result is
\bea
\left(\frac{\partial U^0}{\partial x^{\mu}}\right)_{U^i}&=& - \frac{{U^0}^{2}}{2Q}
\frac{\partial g_{\alpha \beta}}{\partial x^\mu} V^\alpha V^\beta,\qquad \textrm{and}\\
\left(\frac{\partial U^0}{\partial U^{i}}\right)_{x^{\mu}}&=& - \frac{{U^0}}{Q}
(g_{i0} V^0+g_{ik} V^k)
\eea
where
\bea
Q = U^0 (g_{0 0} + g_{0 l} V^l).
\eea
Thus, the required derivatives in equation (\ref{liouville2}) are
\bea
\left(\frac{\partial V^j}{\partial x^\mu}\right)_{U^i}&=&  \frac{U^0}{2Q} V^j
\frac{\partial g_{\alpha \beta}}{\partial x^\mu} V^\alpha V^\beta,\label{DVDX}\\
\left(\frac{\partial V^j}{\partial U^i}\right)_{x^{\mu}}&=&
\;\; \frac{V^{j}}{Q}(g_{0i} + g_{ik} V^k);\qquad\mbox{for $i
\neq j$},\nonumber\\
&&\label{DVDU}\\
\hspace{.81cm}&=&-\frac{1}{Q} (c^2U^{0^{-2}} + \sum_{k \neq i} V^k (g_{0 k}
+ g_{kl} V^l))\qquad\mbox{for $i = j$},\nonumber
\eea
(note that $\left({\partial V^j}/{\partial x^{\mu}}\right)_{U^i}$ differs from the expression
 shown in \cite{res} by a sign).
After some calculations (see Appendix \ref{sec:ap2}) one can verify that, up to order $c^{-2}$,
the Liouville's operator in equation (\ref{liouville2}) can be expressed as
\begin{equation}\label{liouville-final}
\begin{split}
{\cal L}_{v}=&\;{\cal L}^{cl}+{\cal L}^{pn}\\
=&\;\frac{\partial}{\partial t}+v^{i}\frac{\partial}{\partial x^{i}}-
\frac{\partial \phi}{\partial x^{i}}\frac{\partial }{\partial v^{i}}+
\frac{1}{c^{2}}\left(\frac{\mathbf{v}^{2}}{2}-\phi \right)\left(\frac{\partial}{\partial t}
+v^{i}\frac{\partial}{\partial x^{i}}\right)\\
&\;+\frac{1}{c^{2}}\left[4v^{i} v^{j}\frac{\partial \phi}{\partial x^{j}}-
\left(\frac{3\mathbf{v}^{2}}{2}+3\phi\right)\frac{\partial \phi}{\partial x^{i}}+
3v^{i}\frac{\partial \phi}{\partial t}-\frac{\partial \psi}{\partial x^{i}}-\frac{\partial \xi_{i}}{\partial t}
-v^{j}\left(\frac{\partial \xi_{i}}{\partial x^{j}}-\frac{\partial \xi_{j}}{\partial
  x^{i}}\right)\right]\frac{\partial}{\partial v^i}
\end{split}
\end{equation}
where ${\cal L}^{cl}$  is the classical Liouville operator (the first three terms of the r.h.s.) and
${\cal L}^{pn}$ is the corresponding post-Newtonian correction (all terms multiplied by $1/c^{2}$).
Then, equation (\ref{liouville2}) now reads
\begin{equation}
    \left({\cal L}^{cl}+{\cal L}^{pn}\right)f(\mathbf{x},\mathbf{v},t)=0.
\end{equation}
Thus we conclude that the CBE in the 1PN approximation can be splitted into a
 Newtonian contribution and a post-Newtonian one, as was obtained by \cite{res}, but
 now with the corrected version for the operator ${\cal L}^{pn}$.

Similar to the classical case, the 1PN equation (\ref{liouville-final})
can be expressed in various ways (see Appendix \ref{sec:ap}), each of
which is useful in different contexts. First, as a vanishing total derivative,
\be
\frac{d f}{dt}=0,\label{form1}
\ee
meaning that the flow through phase space of the \textit{probability fluid},
 (as seen by an observer moving with the particle)
is incompressible \cite{BT}. Second, in terms of Poisson brackets,
\be
\frac{\partial f}{\partial t}+\{f,H\}=0,\label{form2}
\ee
where $H$ is the 1PN Hamiltonian. Since all integrals of motion must
commute with $H$, it implies that Jeans theorem \cite{jeans} is also valid at 1PN order, i.e. that any
static solution of the CBE depends only on the integrals of
motion of the system, and that any function of the integrals
yields a static solution of the CBE. And third, as a continuity equation,
\begin{equation}
    \frac{\partial f}{\partial t}+\frac{\partial}{\partial\mathbf{w}}(f\mathbf{\dot{w}})=0,\label{liouv-alt2}
\end{equation}
where the set $\mathbf{w}=(\mathbf{q},\mathbf{p})$ is an
arbitrary system of canonical coordinates. This equation states
that the probability is conserved in phase space and, upon the
appropriate integration over the momentum space, leads to the
conservation laws in configuration space, i.e. the conservation
of the energy-momentum tensor at 1PN order. We will refrain from
writing out these results here, since they are not particularly illuminating.

\section{Static Solutions of the Post-Newtonian CBE}\label{sec:staticsol}

For systems in static equilibrium $f$  does not depend
explicitly on time and
 the post-Newtonian potential $\xi_{i}$
vanishes. In consequence, equation (\ref{liouv-final-main})
reduces to:
\be
\left[\left(1+\frac{\mathbf{v}^{2}}{2c^{2}}-\frac{\phi}{c^{2}}\right)v^{i}
\frac{\partial}{\partial x^{i}}-\left(1+\frac32\frac{\mathbf{v}^{2}}{c^{2}}+
\frac{3\phi}{c^{2}}\right)\frac{\partial \phi}{\partial x^{i}}\frac{\partial }{\partial v^{i}}+\frac{4v^{i} v^{j}}{c^2}
\frac{\partial \phi}{\partial x^{i}}\frac{\partial }{\partial v^{j}}-
\frac{1}{c^2}\frac{\partial \psi}{\partial x^{i}}\frac{\partial}{\partial v^{i}}\right]\label{liouville-static}
f(\mathbf{x},\mathbf{v})=0.
\ee
Since the energy (eq. (\ref{energia-main})) is an integral of motion of the system, we say that
any ergodic DF $f(E)$ satisfies (\ref{liouville-static}), due to Jeans theorem.

On the other hand, the gravitational fields $\phi$ and $\psi$ are related to the matter-energy distribution
through the Einstein equations which, in the 1PN approximation, can be written as \cite{WB}
\begin{equation}
    \nabla^{2}\phi=4\pi G~^{0}T^{00},
    \qquad\nabla^{2}\psi=4\pi Gc^2\left(^{2}T^{00}
    +~\!\!^{2}T^{ii}\right)\label{ecs-campo}.
\end{equation}
Here, the convention used is
\begin{equation}
T^{\mu\nu}=~\!^{0}T^{\mu\nu}+~\!\!^{1}T^{\mu\nu}+~\!\!^{2}T^{\mu\nu}+...
\end{equation}
so that the symbol $^{N}T^{\mu\nu}$ refers the
$\mu\nu$-component, of order $(\bar{v}/c)^{N}$, in the expansion of the energy-momentum tensor.
In particular, $^{0}T^{00}$ is the density of rest-mass,
 $^{2}T^{00}$ is the nonrelativistic part of the
energy density and  $^{2}T^{ii}$ (summation over $i$) is the classical kinetic energy density.

Now, in general, the energy-momentum tensor is related to the DF through the equation
\begin{equation}\label{tmunu}
T^{\mu\nu}(x^\lambda)=\frac{1}{c}\int\frac{ U^\mu U^\nu}{U^0}f(x^\lambda,U^i)\sqrt{-g}d^3U,
\end{equation}
so that relations (\ref{liouville-static})-(\ref{ecs-campo}) form a set of
self-consistent equations. For practical purposes it is necessary to expand (\ref{tmunu})
at various orders in $\bar{v}/c$ so, in order to illustrate this idea, we are now going to
focus in the same special case used in \cite{res}, i.e. the post-Newtonian spherical polytropes.
Needless to say, in the future it would be also interesting to consider the case of anisotropic systems,
or axially symmetric systems, specially to develop applications for galactic dynamics.

\subsection{Construction of Post-Newtonian Polytropes}

Now we deal with the 1PN version of polytropic solutions, i.e.
systems characterized with DFs of the form
\begin{equation}\label{df-poli}
\begin{split}
f(E)=&\frac{k_{n}}{2\pi}(-E)^{n-3/2}\quad\mbox{for }E<0,\\
=&\:\:0\quad\quad\quad\quad\;\;\;\:\:\mbox{for }E\geq0,
\end{split}
\end{equation}
where $k_{n}$ is a real constant and $n$ is the index of the polytrope.
The DF can be splitted in two parts: a $c^{0}$ (classical) contribution $f^{(0)}$ and
a $c^{-2}$ (post-Newtonian) contribution $f^{(2)}$. This is possible because the energy can be written as
\begin{equation}\label{energy-split}
    E=E_{cl}+E_{pn}
\end{equation}
where
\begin{equation}
\label{energia-split2}
E_{cl}
=\frac{\mathbf{v}^{2}}{2}+\phi,\qquad
E_{pn}=\frac{3\mathbf{v}^{4}}{8c^2}-\frac{3\mathbf{v}^{2}\phi}{2c^2}+\frac{\phi^2}{2c^2}+\frac{\psi}{c^2}.
\end{equation}
Since we assume that $E_{cl}\gg E_{pn}$, we can write
\begin{equation}
\begin{split}
f&\approx \frac{k_{n}}{2\pi}\left(-E_{cl}\right)^{n-3/2}
\left[1+\left(n-3/2\right)E_{pn}/E_{cl}\right]\\
&=
\frac{k_{n}}{2\pi}\left(-E_{cl}\right)^{n-3/2}-
\frac{k_{n}}{2\pi}\left(n-3/2\right)E_{pn}\left(-E_{cl}\right)^{n-5/2}\\
&=f^{(0)}+f^{(2)}
\end{split}
\end{equation}
In order to obtain $T^{\mu\nu}$ at various orders
of $\bar{v}/c$, we will need the expansion of $U^0$ (see eq. (\ref{Ucero})),
the expansion of the determinant of the metric tensor,
$\sqrt{-g}=1-2\phi/c^2+...$
and also the expansion of $d^3U$, rewritten in terms of $d^3v$.
Remembering that $U^i=U^0v^i/c$ and using (\ref{Ucero}), we have
\begin{equation}
\begin{split}
\frac{\partial U^i}{\partial v^j}=&\;\frac{\partial}{\partial v^j}
\left[\left(1+\frac{\mathbf{v}^{2}}{2c^2}-\frac{\phi}{c^2}\right)v^i\right]\\
=&\;\left(1+\frac{\mathbf{v}^{2}}{2c^2}-\frac{\phi}{c^2}\right)\frac{\partial v^i}{\partial v^j}+
v^i\frac{\partial}{\partial v^j}\left(\frac{\mathbf{v}^{2}}{2c^2}\right)\\
=&\;\left(1+\frac{\mathbf{v}^{2}}{2c^2}-\frac{\phi}{c^2}\right)\delta^i_j+\frac{v^iv_j}{c^2},
\end{split}
\end{equation}
which means that the Jacobian of the transformation is
$|\partial U^i/\partial v^j|=1+5\mathbf{v}^{2}/(2c^2)-3\phi/c^2+...$
and, in consequence,
\begin{equation}
d^3U=\left(1+\frac{5\mathbf{v}^{2}}{2c^2}-\frac{3\phi}{c^2}\right)d^3v=
4\pi\left(1+\frac{5\mathbf{v}^{2}}{2c^2}-\frac{3\phi}{c^2}\right)\mathbf{v}^{2}dv
\end{equation}
(the r.h.s. of the last equation can be implemented only in the case of a DF
depending on the velocity components through $v=(v_iv^i)^{1/2}$). Putting all of this together,
we can now write $T^{00}$ and $T^{ii}$ as:
\begin{equation}\label{T00-gen}
    T^{00}=4\pi\int_{0}^{v_{e}}\left[f^{(0)}+f^{(2)}\right]
    \left(1+\frac{3\mathbf{v}^{2}}{c^{2}}-\frac{6\phi}{c^{2}}\right)\mathbf{v}^{2}dv
\end{equation}
and
\begin{equation}\label{Tij-gen}
    T^{ii}=\frac{4\pi}{c^2} \int_{0}^{v_{e}}\left[f^{(0)}+f^{(2)}\right]
    \left(1+\frac{3\mathbf{v}^{2}}{c^{2}}-\frac{6\phi}{c^{2}}\right)\mathbf{v}^{4}dv.
\end{equation}
The components $T^{0i}=T^{i0}$ vanish due to the distribution of
matter is static (see \cite{WB}). Here $v_{e}$ denotes the escape velocity, i.e.
 the speed at which a particle reaches its maximum value of energy, $E=0$,
 so that it is confined to the distribution of matter.
 Such quantity can be computed from (\ref{energia-main}) and the result is
\begin{equation}\label{vel-escape}
\begin{split}
v_{e}=&\sqrt{-\frac{2 c^2}{3}+2 \phi +\frac{2}{3} \sqrt{c^4-12 c^2 \phi +6 \phi ^2-6 \psi }}\\
\approx &\sqrt{-2\phi }-\frac{1}{c^2}\sqrt{\frac{-\phi^3}{2} }+\frac{1}{c^2}\sqrt{\frac{-\psi^2 }{2 \phi }  }.
\end{split}
\end{equation}
In order to be consistent with the 1PN approximation, one can set $v_{e}=\sqrt{-2\phi}$ due to the fact that
the $c$-dependent terms would lead to the apparition of a factor of
 order $\mathcal{O}(c^{-6})$, even in the case
when the integrand is $f^{(0)}$.
Introducing this value in (\ref{T00-gen})-(\ref{Tij-gen}),
we can obtain explicitly $^{0}{T^{00}}$,
$^{2}{T^{00}}$ and $^{2}{T^{ii}}$ (see Appendix \ref{sec:ap1})
and, after some calculations, we obtain
\bea
    \nabla^{2}\phi&=&\alpha_{n}(-\phi)^{n},\label{eqs-campo-explic1}\\
    \nabla^{2}\psi&=&-n\alpha_{n}(-\phi)^{n-1}\psi+ \beta_{n}(-\phi)^{n+1},\label{eqs-campo-explic2}
\eea
where $\nabla^{2}=(1/r^{2})(d/dr)(r^{2}d/dr)$ and we have introduced the constants
\bea
\alpha_{n}&=&4\sqrt{2}\:\:\pi^{3/2}G k_{n}\frac{\Gamma(n-1/2)}{\Gamma(n+1)},\\
\beta_{n}&=&-2\sqrt{2}\:\:\pi^{3/2}G k_{n}\frac{(n^{2}-2n-63/4)\Gamma(n-1/2)}{\Gamma(n+2)}.\nonumber
\eea
The first equation above (\ref{eqs-campo-explic1})
(which looks like very different from the one derived in \cite{res})
is the  classical field equation for the
Newtonian polytropes \cite{eddi} and has simple exact solutions for the cases
$n=0,1,5$, the latter corresponding to the Plummer's model \cite{schus,plummer}. For other values of $n$, the solution
can not be expressed in terms of elementary functions \cite{BT}.

\subsection{Numerical Solutions of the Field Equations}

In order to perform the numerical solution of the system for any $n$, we implement the following definitions:
\be
\begin{split}
\tilde{r}=&\sqrt{\alpha_{n}(-\phi_{o})^{n-1}}r,\qquad
B_{n}=\frac{(n^{2}-2n-63/4)\phi_{o}^{2}}{2(n+1)\psi_{o}},\\
\phi=&\phi_{o}X,\qquad
\psi=\psi_{o}Y,
\end{split}
\ee
where $\phi_{o}$ and $\psi_{o}$ are the Newtonian and post-newtonian gravitational potentials
at the center of the configuration, respectively. Then, relations (\ref{eqs-campo-explic1})-(\ref{eqs-campo-explic2})
become
\bea
\frac{1}{\tilde{r}^{2}}\frac{d}{d \tilde{r}}\left(\tilde{r}^{2}\frac{d X}{d \tilde{r}}\right)&=&-X^{n},\\
\frac{1}{\tilde{r}^{2}}\frac{d}{d \tilde{r}}
\left(\tilde{r}^{2}\frac{d Y}{d \tilde{r}}\right)&=&-n X^{n-1}Y-B_{n}X^{n+1}.
\eea
Since we assume that the gravitational potentials reach critical values at the center
of the configuration, we have to impose the initial conditions
\begin{equation}\label{init-conditions}
    X(0)=Y(0)=1,\qquad X'(0)=Y'(0)=0,
\end{equation}
where the prime denotes differentiation with respect to the scaling radius $\tilde{r}$. We use
a fourth-order Runge-Kutta method to find the numerical solutions for different values of the ratios $\phi_{o}/c^{2}$ and assuming
$\psi_{o}\sim-\phi_{o}^{2}$. In a neutron star, for example,
$\phi_{o}/c^{2}$ can vary between $-1$ and $-0.1$, approximately.

\begin{figure*}
  $$
\begin{array}{cc}
  \epsfig{width=3.0in,file=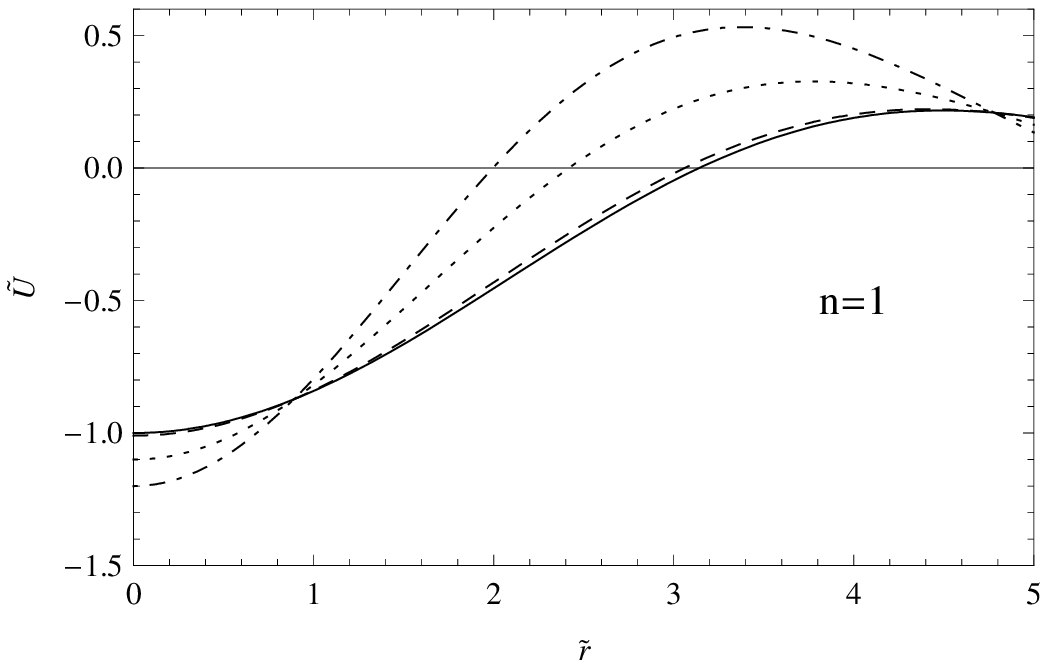} & \epsfig{width=3.0in,file=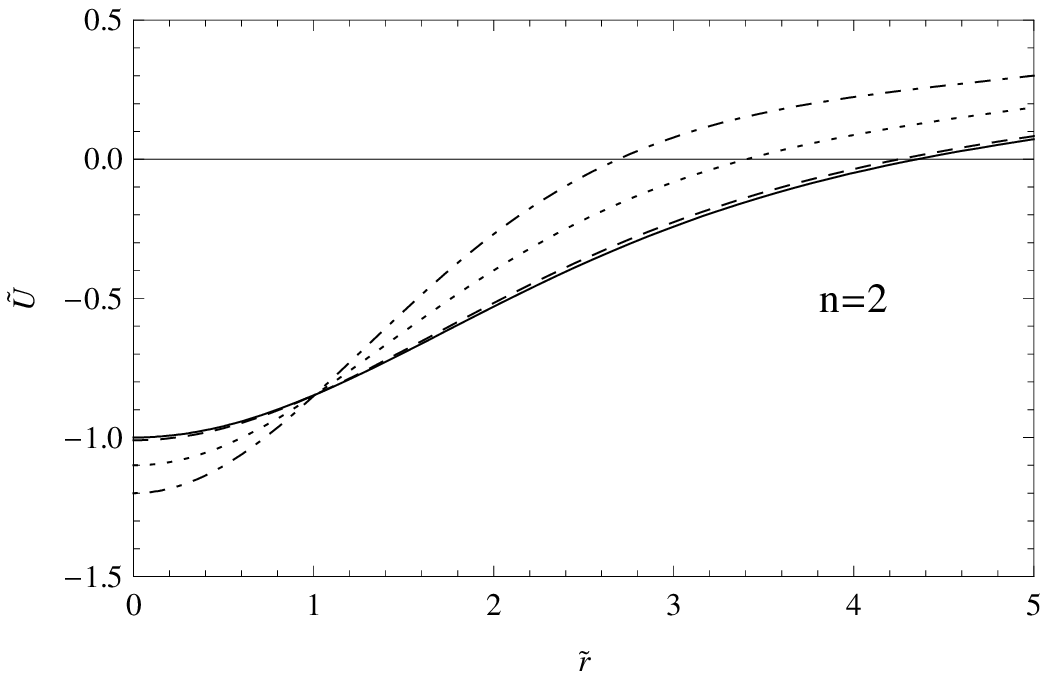} \\
  \epsfig{width=3.0in,file=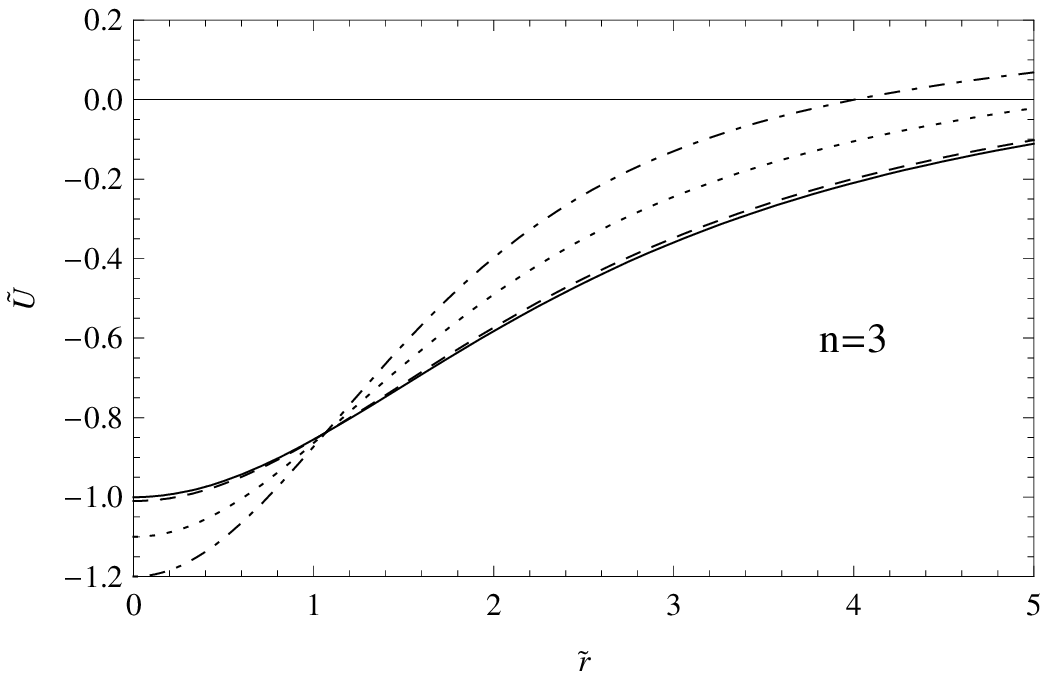} & \epsfig{width=3.0in,file=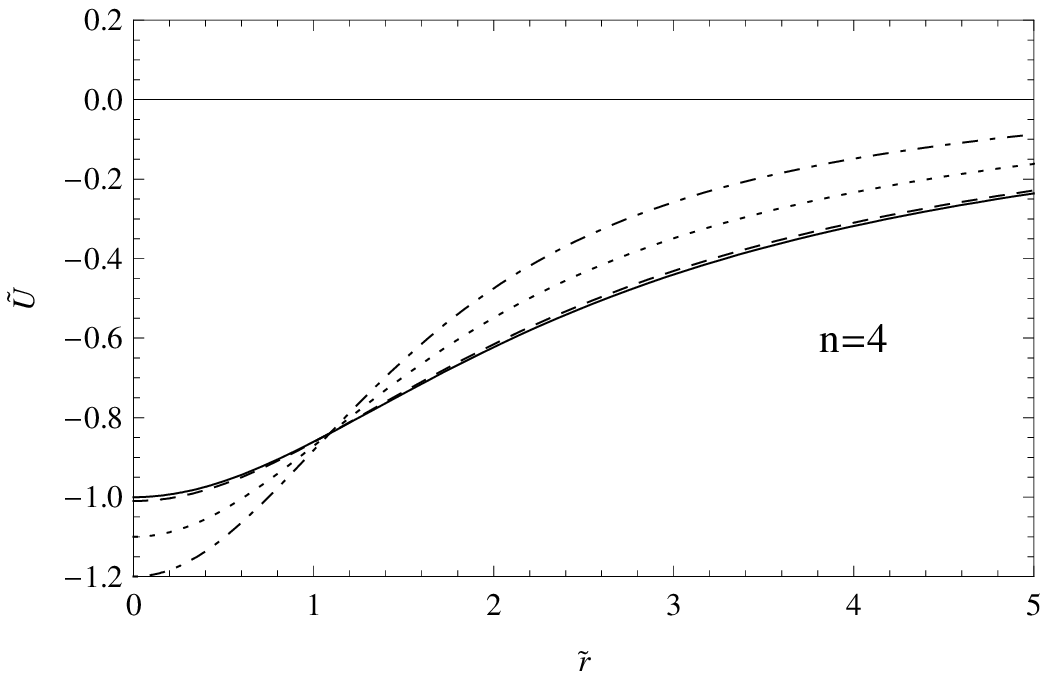}\\
  \epsfig{width=3.0in,file=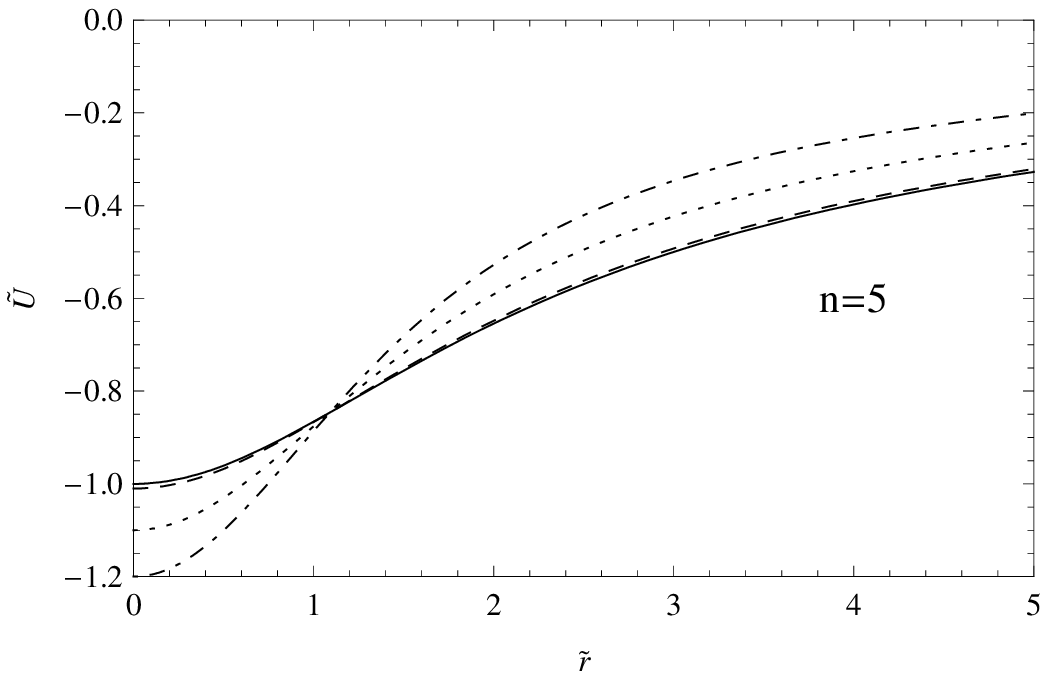} & \epsfig{width=3.0in,file=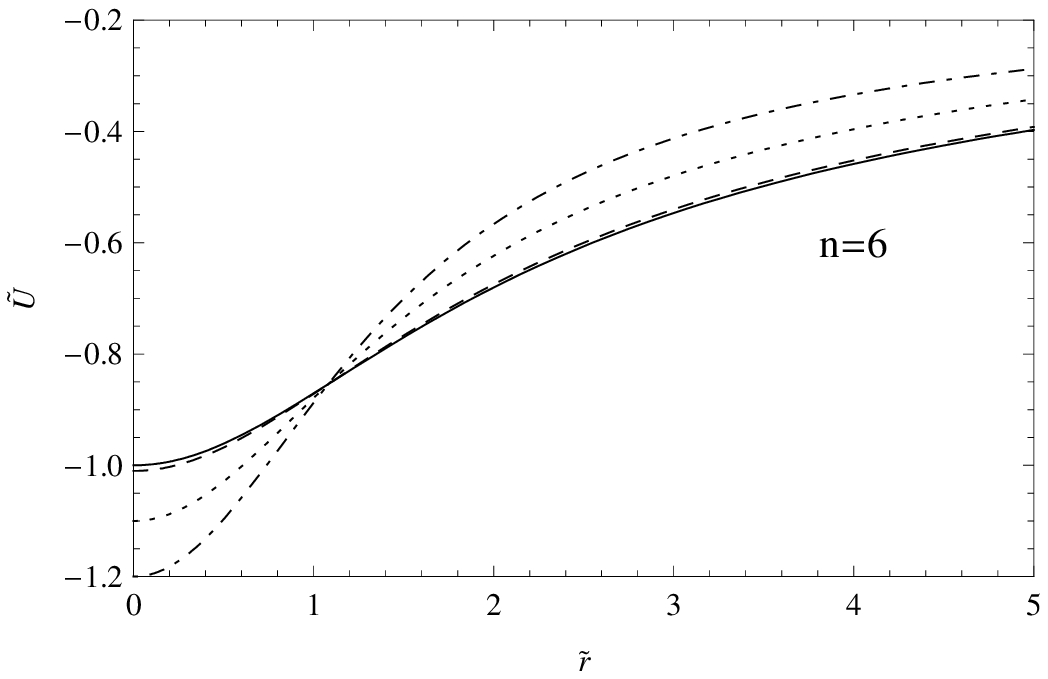}
\end{array}
$$
   \caption{We show the gravitational potential energy $\tilde{U}$ (given by eq.(\ref{U}))
   for Newtonian (continuous line) and
   Post-Newtonian polytropes with index
   $n=1,..,6$. In each illustration we assume $\psi_{o}\sim\phi_{o}^{2}$ and choose
   $\phi_{o}/c^{2}$ equal to $-0.02$ (dashed line), $-0.2$ (dotted line) and $-0.4$ (dash-dotted line).}
              \label{figure}%
    \end{figure*}

In Figure \ref{figure} we plot the gravitational potential energy of a test particle with velocity $v=0$
(scaled with respect to $\phi_{o}$), given by
\begin{equation}\label{U}
    \tilde{U}=-U/\phi_{o}=-X-\frac{\phi_{o}}{2c^{2}}X^{2}-\frac{\psi_{o}}{c^{2}\phi_{o}}Y,
\end{equation}
for the case of Newtonian polytropes and post-Newtonian polytropes.
We note that for $\phi_{o}/c^{2}\sim -0.02$ or smaller,
the post-Newtonian corrections are not significant,
 while for values $\phi_{o}/c^{2}\sim -0.2$ or larger they become important.

\subsection{Post-Newtonian corrections to the rotation curves and mass densities\label{sec:vc-densmass}}

\begin{figure*}
  $$
\begin{array}{cc}
  \epsfig{width=3.0in,file=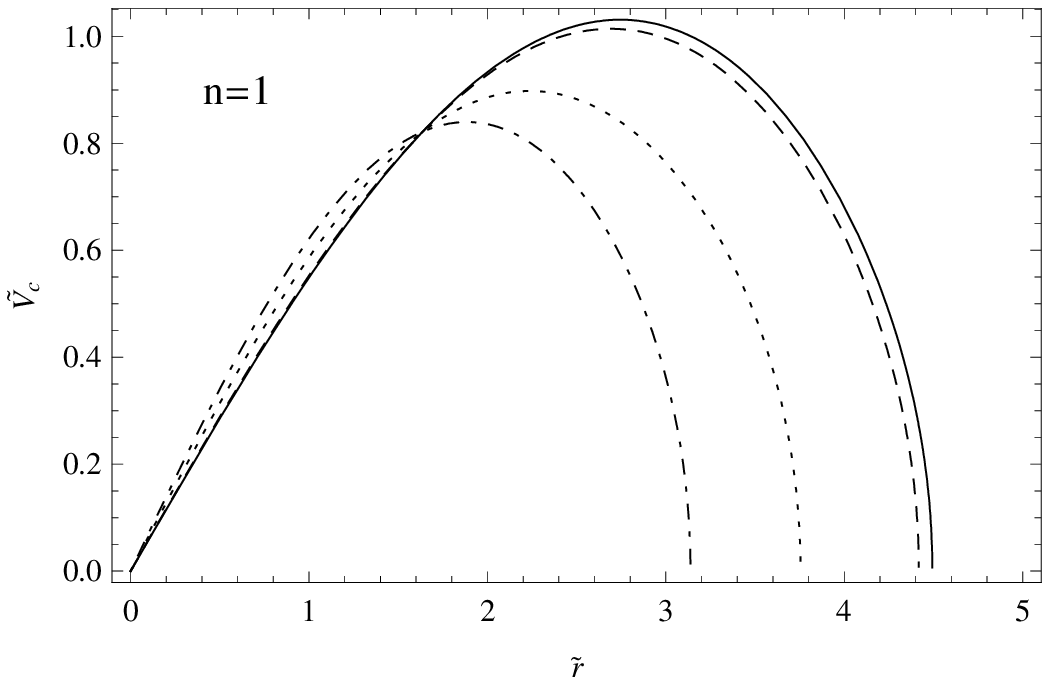} & \epsfig{width=3.0in,file=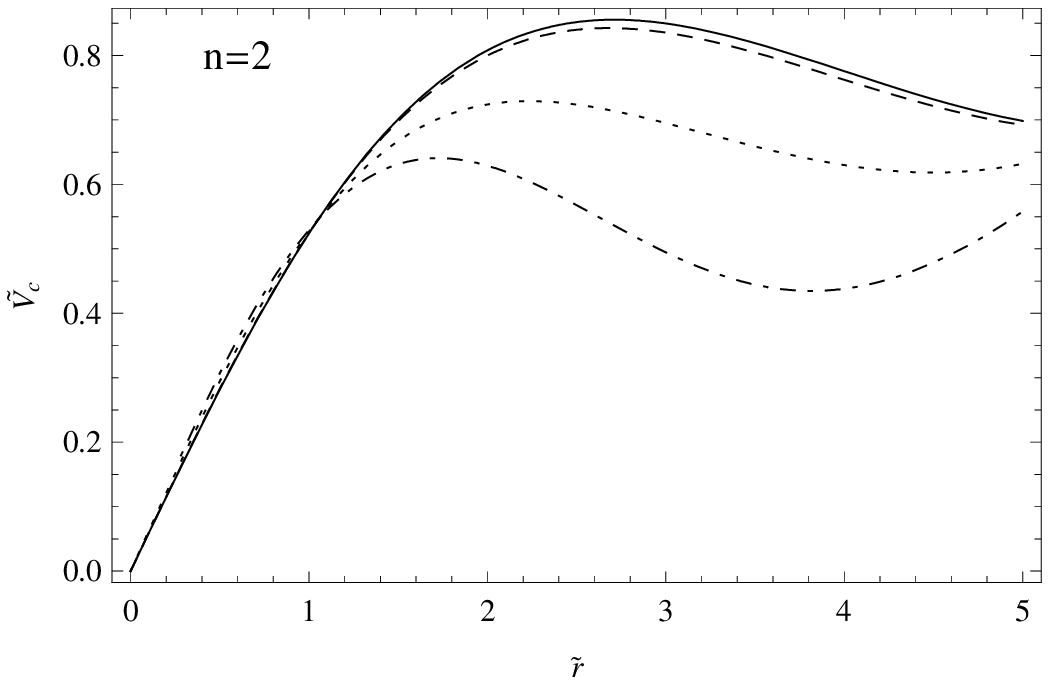} \\
  \epsfig{width=3.0in,file=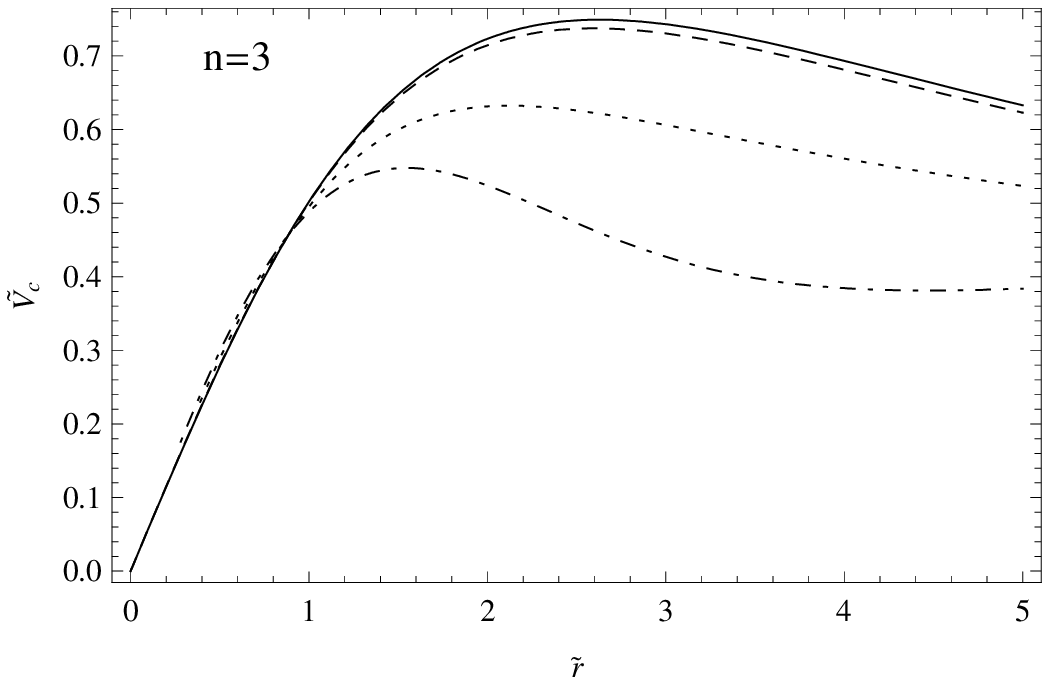} & \epsfig{width=3.0in,file=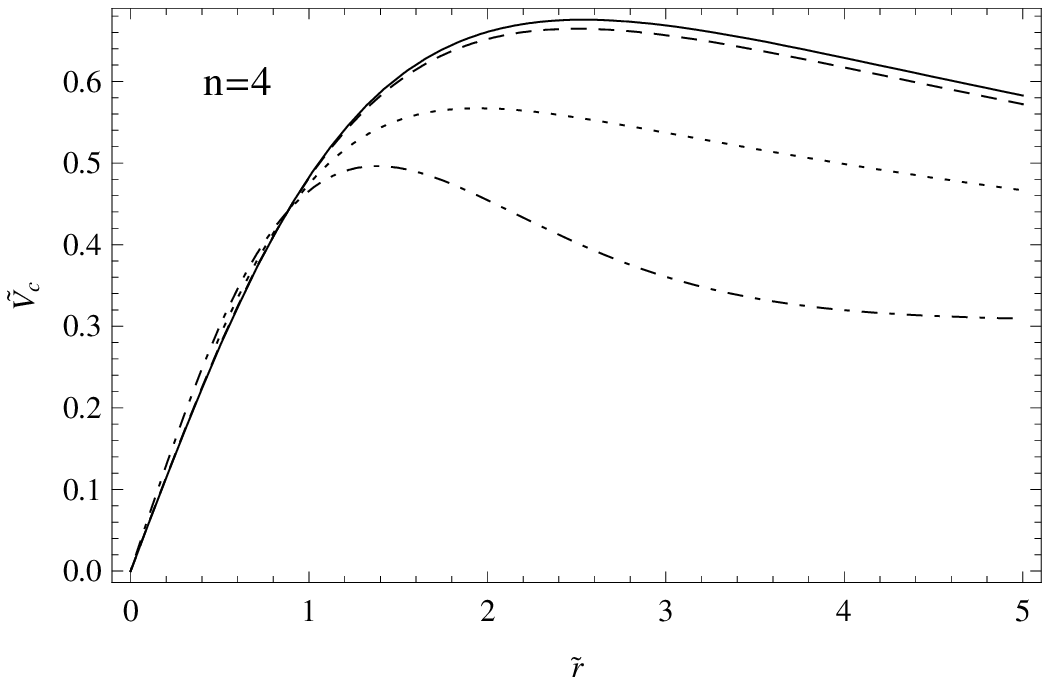}\\
  \epsfig{width=3.0in,file=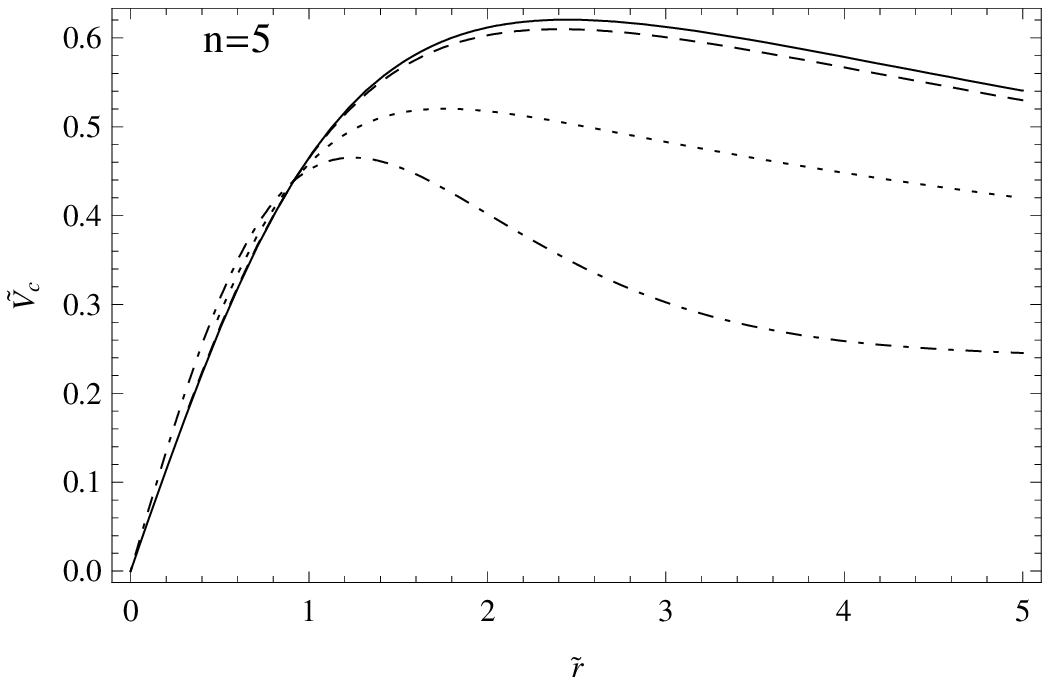} & \epsfig{width=3.0in,file=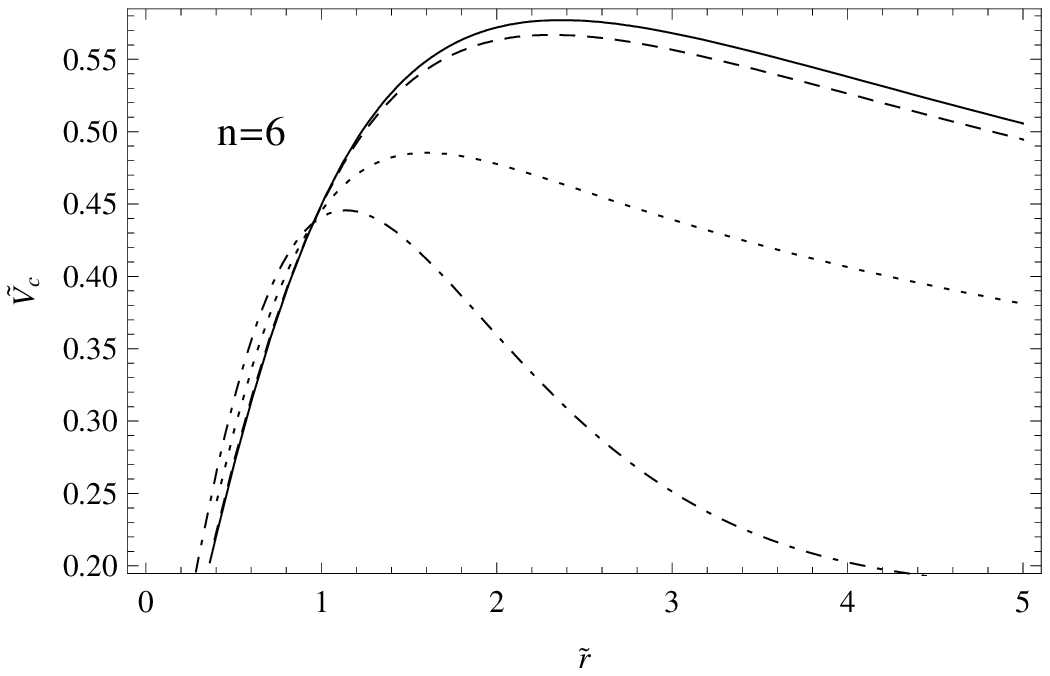}
\end{array}
$$
   \caption{We show the circular velocity $\tilde{V_{c}}$
   for Newtonian (continuous line) and
   Post-Newtonian polytropes with index
   $n=1,..,6$. In each illustration we assume $\psi_{o}\sim\phi_{o}^{2}$ and choose
   $\phi_{o}/c^{2}$ equal to $-0.02$ (dashed line), $-0.2$ (dotted line) and $-0.4$ (dash-dotted line).}
              \label{figure:vc}%
    \end{figure*}

In this subsection we study the fundamental equations
describing the circular motion of test particles, in
order to investigate the corrections introduced by the relativistic effects on the rotation curves.
In spherical coordinates $(r,\theta,\varphi)$, the circular orbits in the equatorial plane can be obtained from
(\ref{1PN-eq-motion}), by setting $\theta=\pi/2$, $\partial \phi/\partial t=0$ and
$\partial \phi/\partial\varphi=\partial \psi/\partial\varphi=0$.
They must satisfy the conditions $\dot{r}=\dot{z}=0$,
$\ddot{r}=\ddot{z}=0$ and $z=0$, so that the equation of motion reduces to
\be
r\dot{\varphi}^{2}\left[1+\frac{r}{c^{2}}\frac{\partial\phi}{\partial r}\right]=
\frac{\partial}{\partial r}\left[\phi+\frac{2\phi^{2}+\psi}{c^{2}}\right].
\ee
This can be used to derive an expression for the circular velocity
$v_{\varphi}=r\dot{\varphi}$, as a function of the radius $r$. The result reads:
\begin{equation}
v_{\varphi}=\left.
\sqrt{r\frac{\partial\phi}{\partial r}\left(1+\frac{4\phi}{c^{2}}-\frac{r}{c^{2}}
\frac{\partial\phi}{\partial r}\right)+\frac{r}{c^{2}}\frac{\partial\psi}{\partial r}}
\right|_{z=0}. \label{circular-velocity}
\end{equation}
Note that in the limit $c\rightarrow\infty$, the above expression reduces to the
usual relation derived in Newtonian theory: $v_{\varphi}=\sqrt{R\partial\phi/\partial R}$.
Perhaps, the most important difference between such relation and (\ref{circular-velocity})
is that, in the Newtonian case, the radical is linear in $\phi$ and its derivatives, whereas
in the 1PN case, it depends on non linear terms involving $\phi$, $\psi$ and derivatives. This
non linear dependence may result significant in some cases and its effects can be observed in
the rotation curves. In Figure \ref{figure:vc} we show the rotation curves corresponding to the 1PN corrected models.
As we can see, in some cases the 1PN corrections are significant while, in other cases, they are practically negligible.

\begin{figure*}
  $$
\begin{array}{cc}
  \epsfig{width=3.0in,file=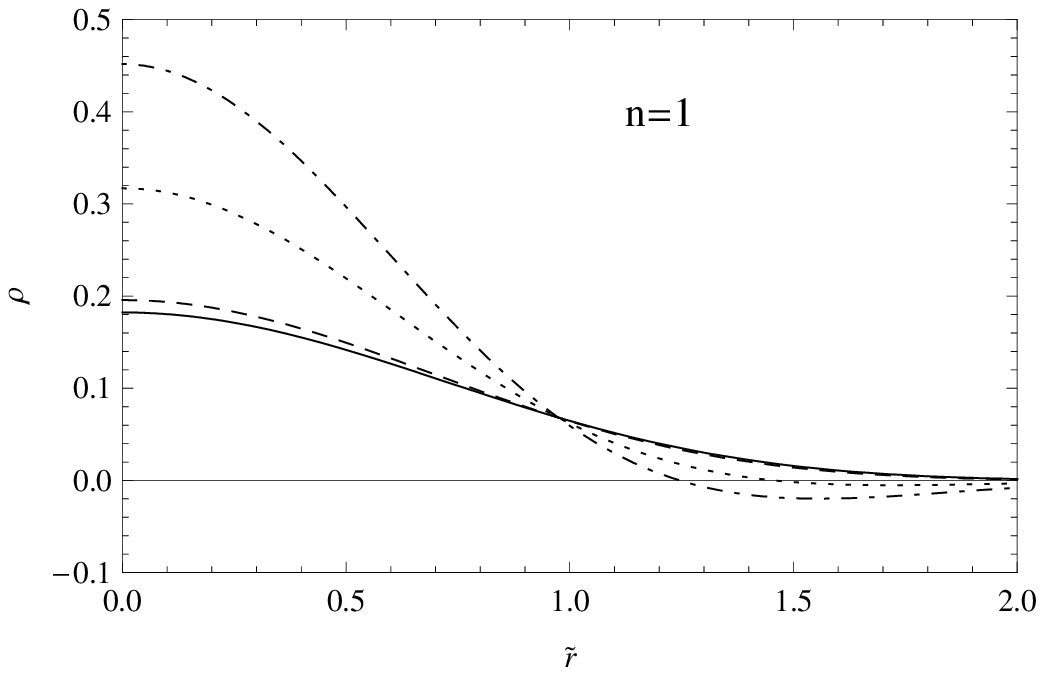} & \epsfig{width=3.0in,file=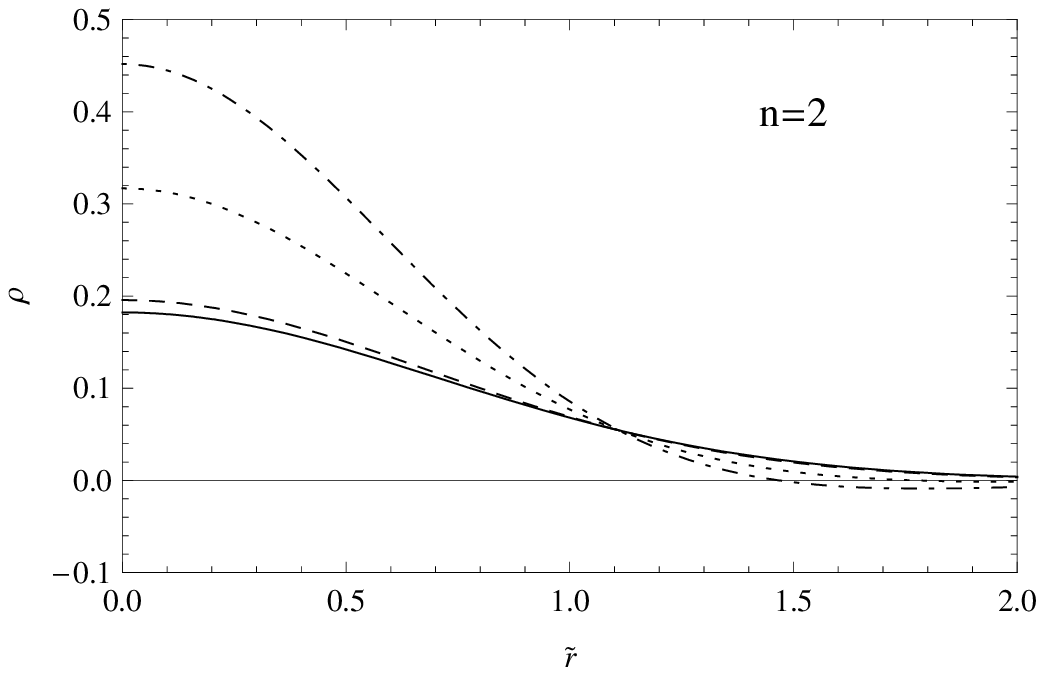} \\
  \epsfig{width=3.0in,file=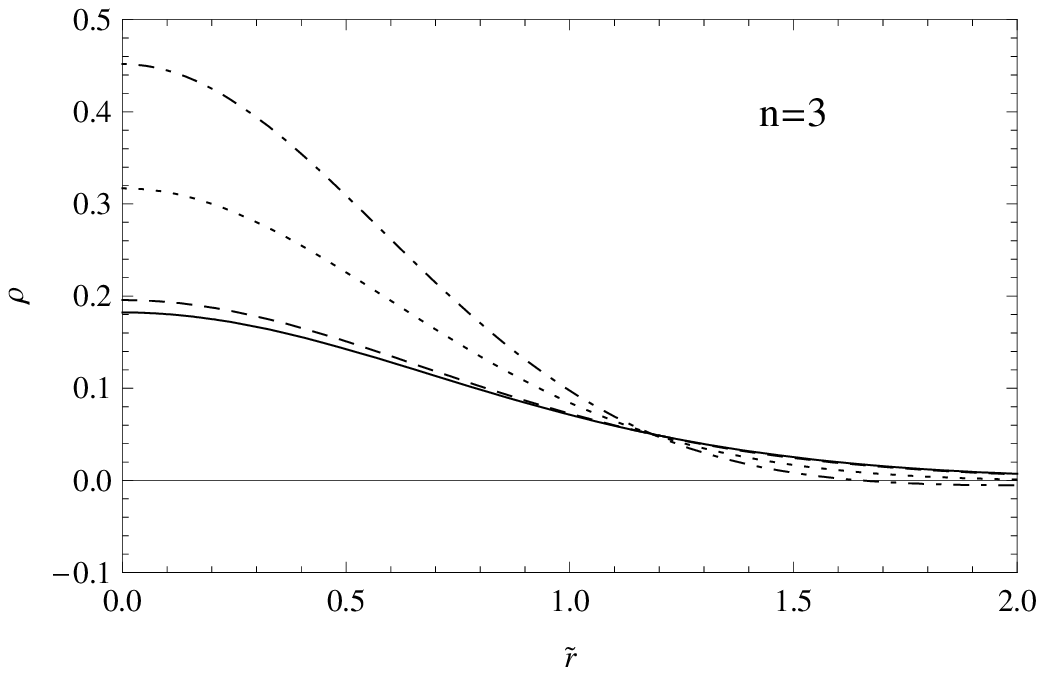} & \epsfig{width=3.0in,file=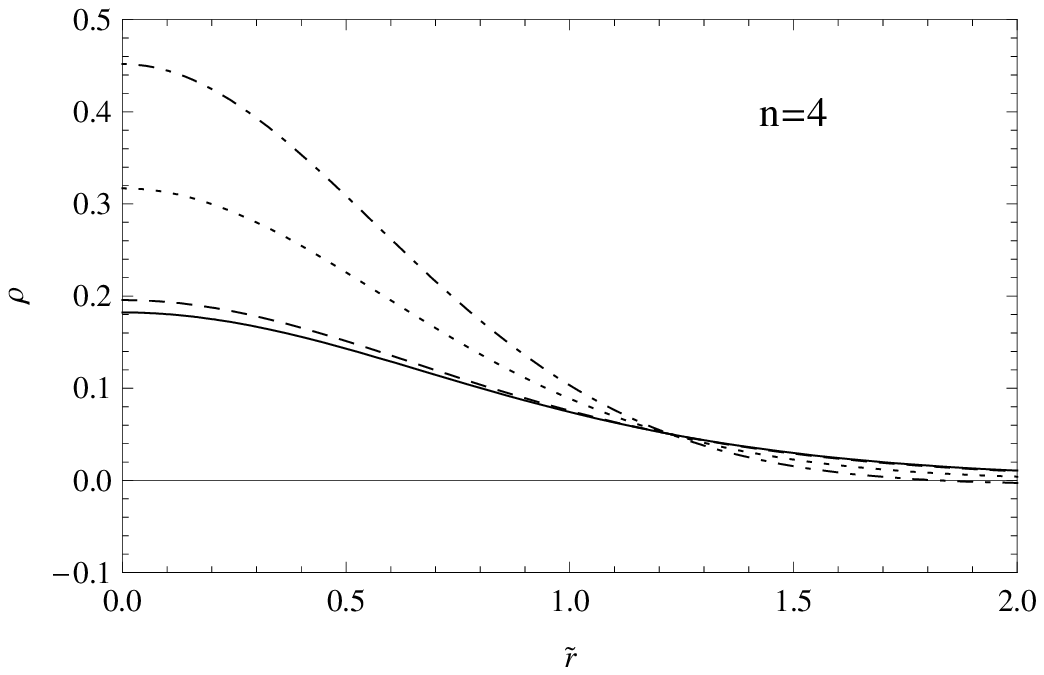}\\
  \epsfig{width=3.0in,file=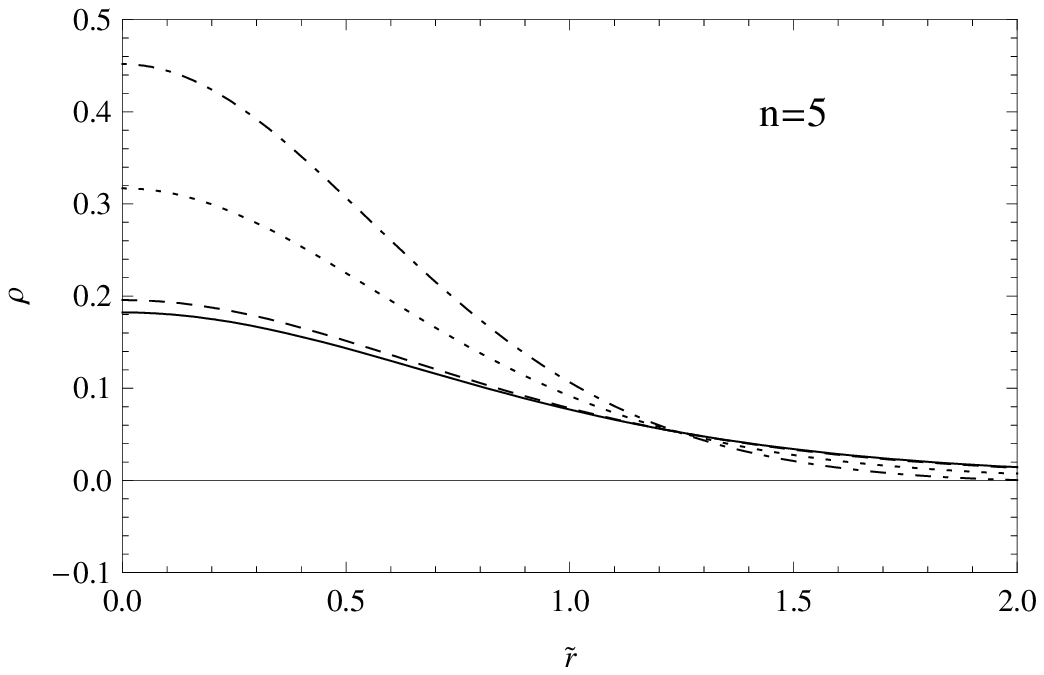} & \epsfig{width=3.0in,file=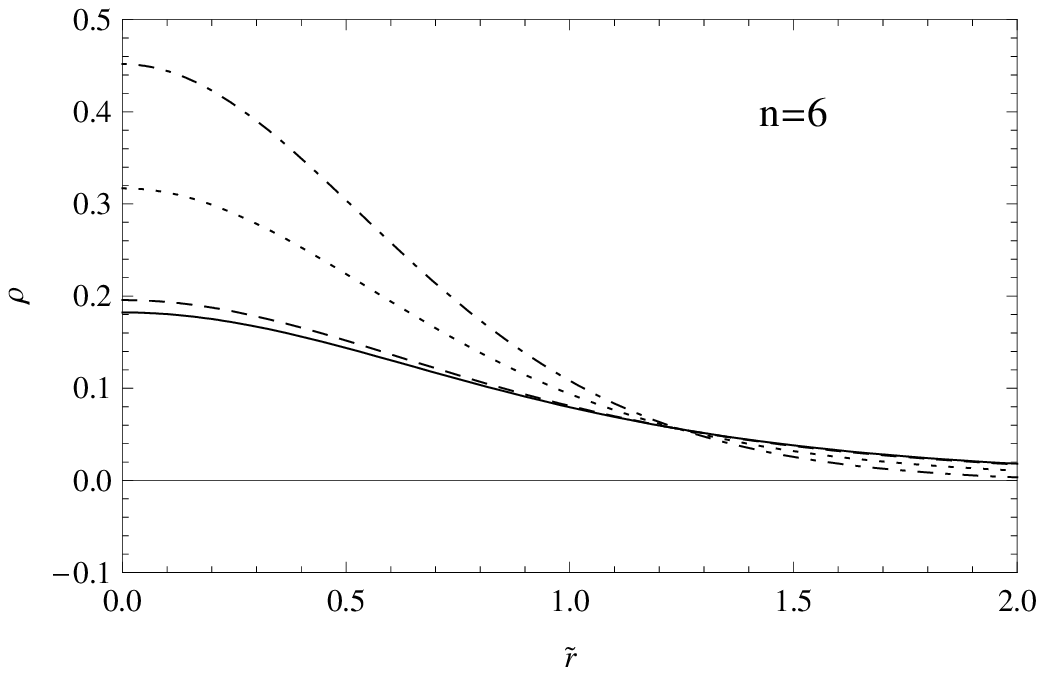}
\end{array}
$$
   \caption{We plot the mass density $\rho$, scaled with respect $k_{n}\phi_{o}^{6}$,
   for Newtonian (continuous line) and
   Post-Newtonian polytropes with index
   $n=1,..,6$. In each illustration we assume $\psi_{o}\sim\phi_{o}^{2}$ and choose
   $\phi_{o}/c^{2}$ equal to $-0.02$ (dashed line), $-0.2$ (dotted line) and $-0.4$ (dash-dotted line).}
              \label{figure:dens}%
    \end{figure*}

\begin{figure*}
  $$
\begin{array}{cc}
  \epsfig{width=3.0in,file=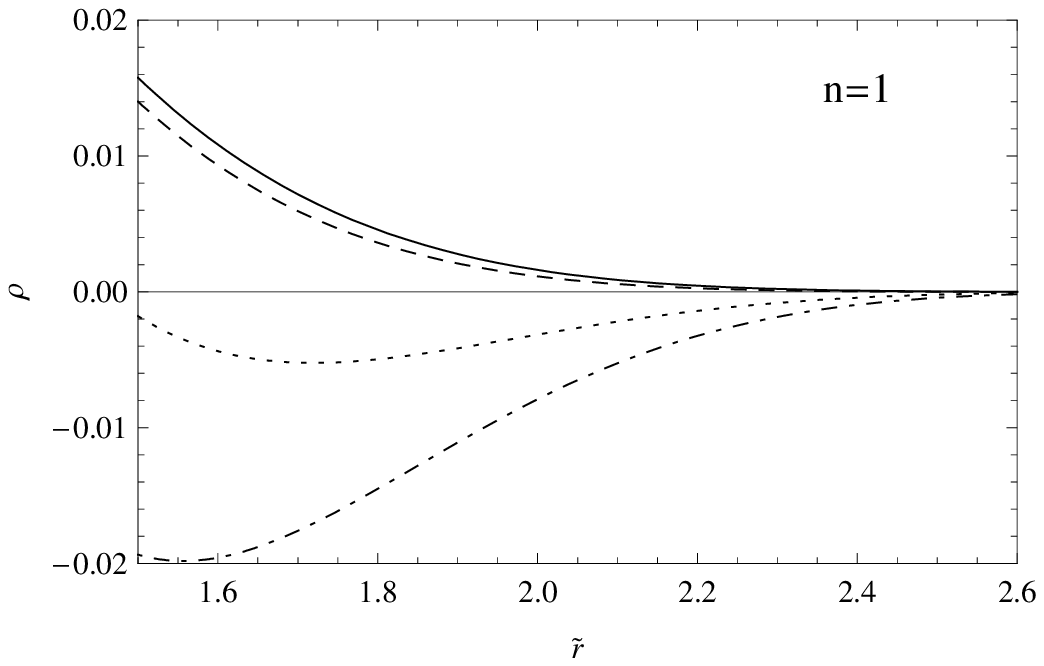} & \epsfig{width=3.0in,file=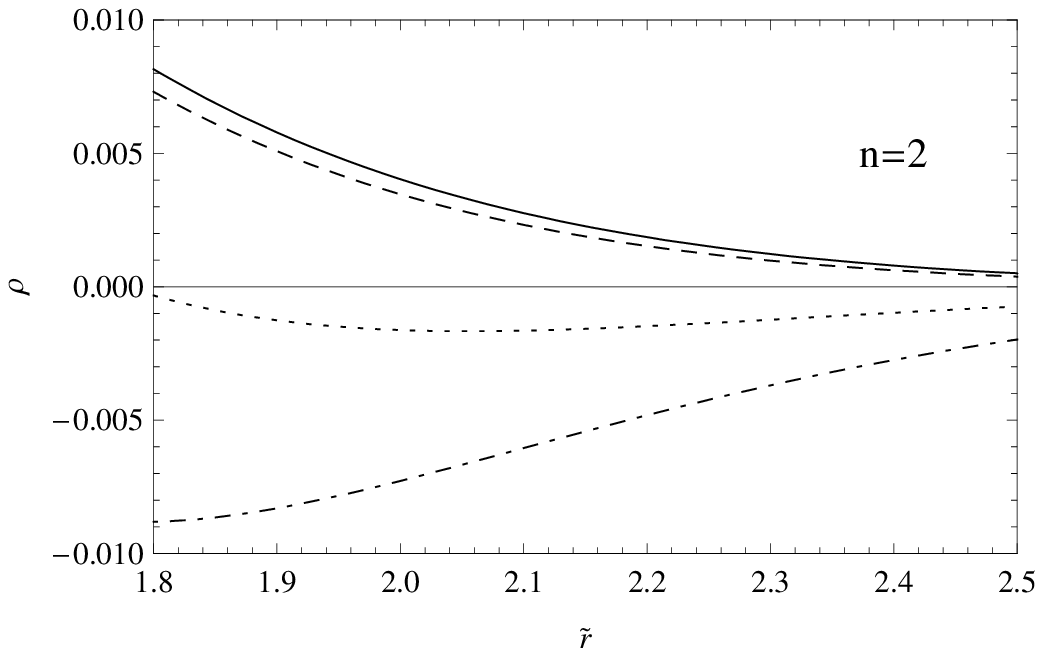} \\
  \epsfig{width=3.0in,file=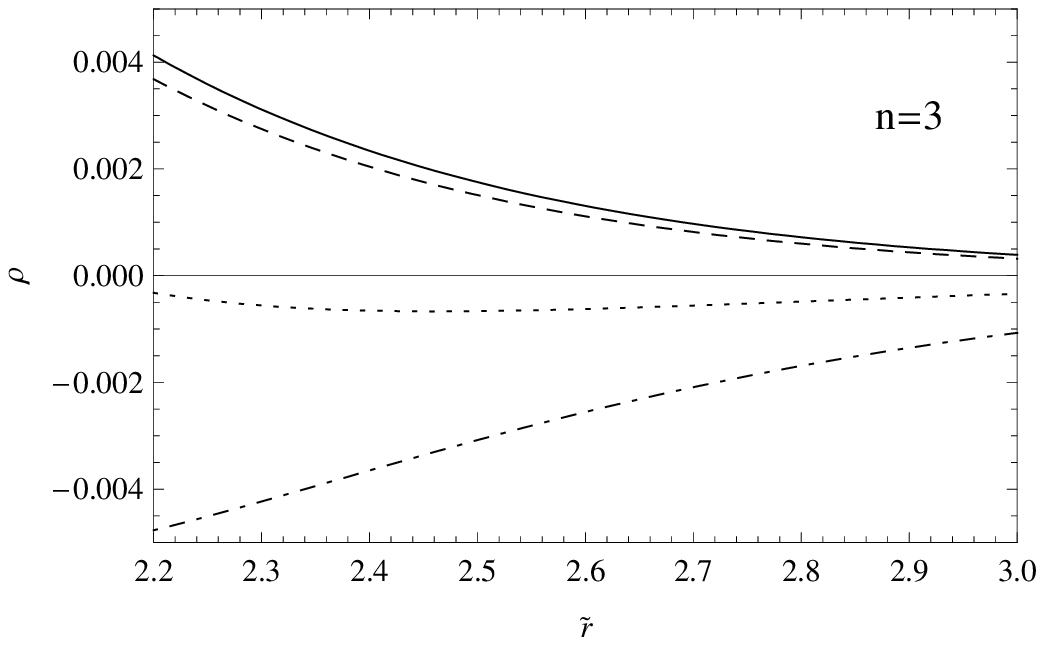} & \epsfig{width=3.0in,file=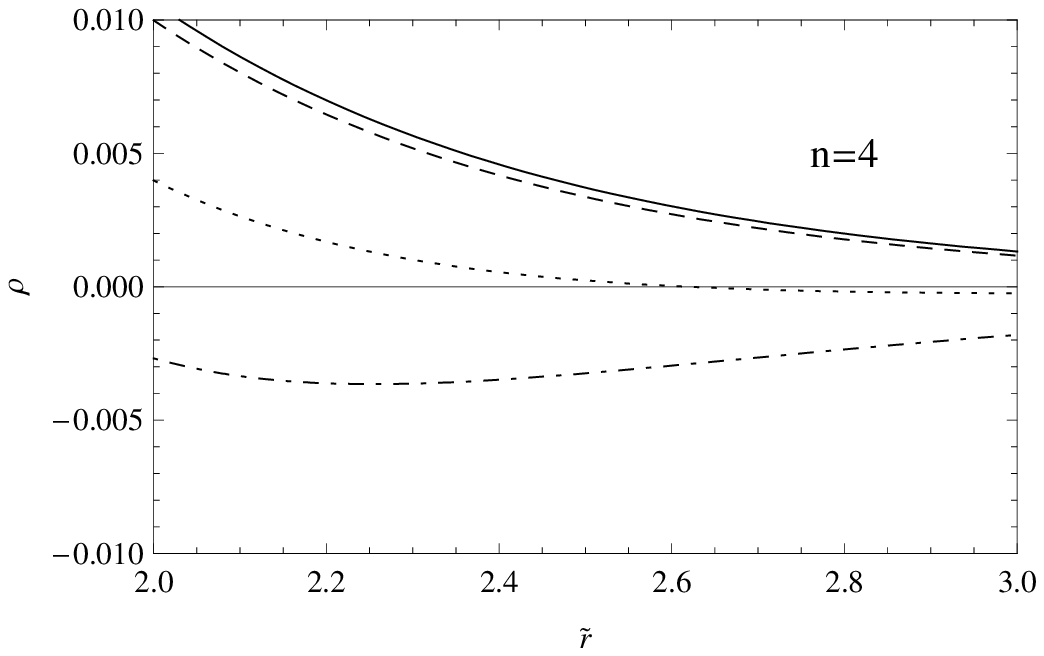}\\
  \epsfig{width=3.0in,file=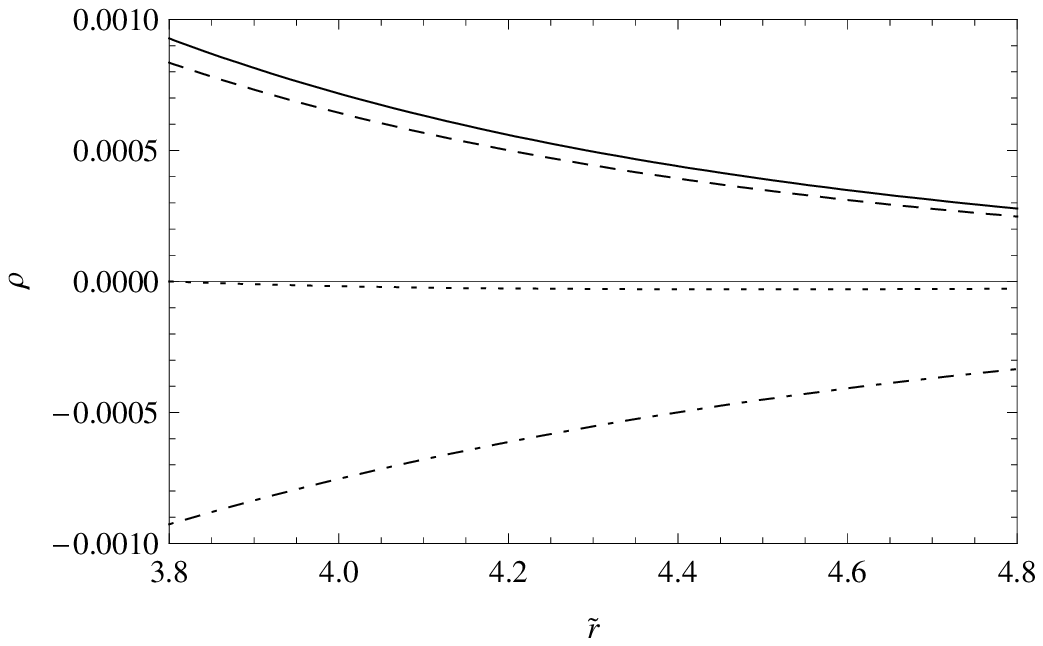} & \epsfig{width=3.0in,file=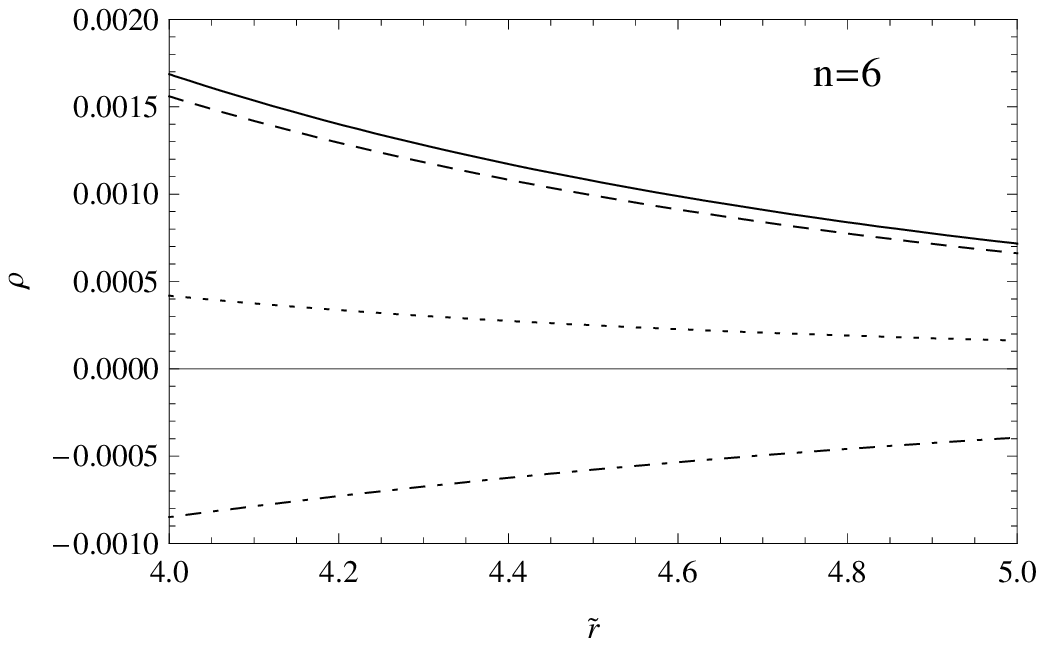}
\end{array}
$$
   \caption{Details of figure  \ref{figure:dens}
   for Newtonian (continuous line) and
   Post-Newtonian polytropes with index
   $n=1,..,6$. In each illustration we assume $\psi_{o}\sim\phi_{o}^{2}$ and choose
   $\phi_{o}/c^{2}$ equal to $-0.02$ (dashed line), $-0.2$ (dotted line) and $-0.4$ (dash-dotted line).}
              \label{figure:dens-a}%
    \end{figure*}

Now, in order to examine if the 1PN corrections provide an adequate physical description for the energy-mass
distribution, we plot the density $\rho$. The contribution of $^{0}T^{00}$ plus
$^{2}T^{00}$ (see Appendix \ref{sec:ap1} for details) as a function of $r$ is shown in Figure
\ref{figure:dens} (here $\rho$ is the mass density divided by $k_{n}\phi_{o}^{6}$, in each case).
We find that $\rho$ is a positive-valued function with a maximum in the center and
a minimum at $r\rightarrow\infty$, for the following situations: (i) polytropes $n=1,\ldots,6$ with $\phi_{o}/c^{2}=-0.02$;
(ii) polytrope $n=6$  with $\phi_{o}/c^{2}=-0.2$. When such ratio reaches the value $-0.4$, the mass density becomes negative
for certain values of $r$, far from the center (see Figure \ref{figure:dens-a}), thus representing a non-physical situation with tachyonic matter.

\section{Concluding Remarks \label{sec:discusion}}

We have obtained the 1PN version of the collisionless Boltzmann equation for a self-gravitating gas
of identical particles, which is consistent with the microscopic equations of motion. This
can be shown by checking that the integrals of motion derived from the microscopic dynamics,
are static solutions of (\ref{liouv-final-main}), in agreement with the statistical description of the system.
Such relation leads to the equations of the post-Newtonian hydrodynamics, derived previously by Chandrasekar \cite{chan}, and implies automatically that the macroscopic stress-energy tensor is conserved if one takes the corresponding integrals over the phase space.

The interpretation of the CBE derived here follows the same logical arguments as in the classical case: (i) we can think of it as a continuity equation of the probability fluid, which is just a statement that comes from the probability conservation in phase space. (ii) The fact that we are allowed to rewrite it as a vanishing total derivative means that the flow of this probability fluid, as seen by a comoving observer, is incompressible. (iii) The structure of the equation in terms of poisson brackets reveals that Jeans theorem is also valid at 1PN order, i.e. that any
static solution of the CBE depends only on the integrals of motion of the system, and that any function of the integrals yields a static solution of the CBE. We suggest some potential relevance of our findings for the study of nuclear cores in galactic dynamics and other astrophysical systems with high enough energy densities such that the relativistic effects become important.

As a first step towards the developing of such astrophysical applications, we considered the case of post-Newtonian polytropes, thus providing the corrected version of the solutions found in \cite{res}. We proceeded to do it numerically because, even in the Newtonian case, it is not possible to find analytical solutions for arbitrary polytropic index. Now, in order to compare the behavior of Newtonian and 1PN solutions, we chose some values for the dimensionless parameters $\phi_{o}/c^{2}$ and $\psi/\phi_{o}^{2}$, so that they can be associated
to relativistic stellar systems. For example, for $\phi_{o}/c^{2}=-0.02$ and $\psi/\phi_{o}^{2}=-1$,
we found that the Newtonian and post-Newtonian behavior are very similar, although some perceptible differences
can be seen in the gravitational energy, rotation curves and  mass densities. In particular, we note that
that the 1PN corrections to the mass densities are more significant near the center of the configurations
and, in contrast, the 1PN corrections to the circular velocities are greater far from the center. In general,
the 1PN values of $\tilde{v}_{c}$ are smaller that the Newtonian values. When we chose $\phi_{o}/c^{2}=-0.2$
or $\phi_{o}/c^{2}=-0.4$, we found that the relativistic contributions are more relevant. However, in some cases the 1PN models are unphysical for these parameters (meaning that the mass density becomes negative for some values of $r$), depending on the polytropic index $n$. We also note
the fact that, the larger are the corrections to the central mass density, the larger are
the corrections to the circular velocity far from the center.

Last but not least, it is worth to point out that it is also possible to derive
 an exact solution for the case $n=5$, i.e. the 1PN version of
 Plummer's model, which is the simplest polytrope with physical relevance.
  It will be presented in a subsequent paper with further applications.

\section*{Acknowledgements}
J.R.-C. is grateful to FAPESP for financial support and to Prof. P. S. Letelier for stimulating discussions.

\appendix

\section{Derivation of relation (\ref{liouville-final}) \label{sec:ap2}}

By introducing (\ref{DVDX}) and (\ref{DVDU}) in the equation
(\ref{liouville2}), the Liouville's operator takes the form
\be
\begin{split}\label{liouville3}
{\cal L}_{U}=&\;U^{0} V^\mu\frac{\partial}{\partial x^\mu}+ \frac{(U^{0})^{2}V^{\mu}}{2Q}
\frac{\partial g_{\alpha \beta}}{\partial x^{\mu}}V^{\alpha}V^{\beta}V^{j}
\frac{\partial}{\partial V^j}-\frac{(U^{0})^{2}}{Q}\Gamma_{\mu\nu}^i V^\mu
V^\nu\sum_{j\neq i}\frac{V^{j}}{Q}(g_{0i} + g_{ik} V^k)\frac{\partial}{\partial V^j}\\
&\;+\frac{(U^{0})^{2}}{Q}\Gamma_{\mu\nu}^i V^\mu
V^\nu\left[c^{2}(U^{0})^{-2}+\sum_{k \neq i} V^k (g_{0 k}
+ g_{kl} V^l)\right]\frac{\partial}{\partial V^i}.
\end{split}
\ee
The four terms in the r.h.s., up to order $c^{-2}$, can be written as
\bea
\label{parte1}
U^{0} V^\mu\frac{\partial}{\partial x^\mu}=\left(1+\frac{\mathbf{v}^{2}}{2c^{2}}-
\frac{\phi}{c^{2}}\right)\frac{\partial}{\partial t} +\left(1+\frac{\mathbf{v}^{2}}{2c^{2}}-
\frac{\phi}{c^{2}}\right)v^{i}\frac{\partial}{\partial x^{i}},\nonumber\\
\eea
\bea
\frac{(U^{0})^{2}V^{\mu}}{2Q}
\frac{\partial g_{\alpha \beta}}{\partial x^{\mu}}V^{\alpha}V^{\beta}V^{j}
\frac{\partial}{\partial V^j}= \frac{v^{i}}{c^{2}}\frac{\partial \phi}{\partial t}
\frac{\partial}{\partial v^i}+\frac{v^{i}v^{j}}{c^{2}}
\frac{\partial \phi}{\partial x^{i}}\frac{\partial}{\partial v^j},\nonumber\\
\eea
\bea
-\frac{(U^{0})^{2}}{Q}\Gamma_{\mu\nu}^i V^\mu
V^\nu\sum_{j\neq i}\frac{V^{j}}{Q}(g_{0i} + g_{ik} V^k)\frac{\partial}{\partial V^j}
=\frac{v^{i}v^{j}}{c^{2}}
\frac{\partial \phi}{\partial x^{i}}\frac{\partial}{\partial v^j}-\frac{(v^{i})^{2}}{c^{2}}
\frac{\partial \phi}{\partial x^{i}}\frac{\partial}{\partial v^i},
\eea
\bea
\label{parte4}
\frac{(U^{0})^{2}}{Q} &\Gamma_{\mu\nu}^i& V^\mu
V^\nu\Big[c^{2}(U^{0})^{-2}+\sum_{k \neq i} V^k (g_{0 k}
+ g_{kl} V^l)\Big]\frac{\partial}{\partial V^i}\nonumber\\
&=&
-\left(1+\frac{3}{2}\frac{\mathbf{v}^{2}}{c^{2}}+\frac{3\phi}{c^{2}}\right)
\frac{\partial \phi}{\partial x^{i}}\frac{\partial}{\partial v^i}+\frac{(v^{i})^{2}}{c^{2}}\frac{\partial \phi}{\partial x^{i}}\frac{\partial}{\partial v^i}
+\frac{2v^{i}v^{j}}{c^{2}}\frac{\partial \phi}{\partial x^{i}}\frac{\partial}{\partial v^j}
\nonumber \\
&& +\frac{2v^{i}}{c^{2}}\frac{\partial \phi}{\partial t}
\frac{\partial}{\partial v^i}-\frac{1}{c^{2}}\frac{\partial \psi}{\partial x^{i}}
\frac{\partial}{\partial v^i}-\frac{1}{c^{2}}\frac{\partial \xi_{i}}{\partial t}\frac{\partial}{\partial v^i}
\nonumber \\&& -\frac{v^{j}}{c^{2}}\left(\frac{\partial \xi_{i}}{\partial x^{j}}-
\frac{\partial \xi_{j}}{\partial x^{i}}\right)\frac{\partial}{\partial v^i}.
\eea
In the derivation of (\ref{parte1})-(\ref{parte4})
we have used the approximations
\bea
\frac{U^{0}}{c}\approx 1-\frac{\phi}{c^2}+\frac{\mathbf{v}^{2}}{2c^{2}},\nonumber\\
\frac{(U^{0})^{2}}{c Q} \approx -1+\frac{3\phi}{c^{2}}-\frac{\mathbf{v}^{2}}{2c^{2}}, \\
c^2 (U^{0})^{-2}=1+\frac{2\phi}{c^2}-\frac{\mathbf{v}^{2}}{c^2}.\nonumber
\eea
Finally, replacing (\ref{parte1})-(\ref{parte4}) in
(\ref{liouville3}) we obtain the final form of the Liouville's
operator, eq. (\ref{liouville-final}).

\section{Alternative expressions for the CBE\label{sec:ap}}
In order to express the CBE as a total derivative,
let's start by writing explicitly the various orders of the DF, i.e.
\be
{\cal L}_v f={\cal L}^{cl}f^{(0)}+({\cal L}^{cl}f^{(2)}+{\cal L}^{pn}f^{(0)})=0,
\ee
where
\be
{\cal L}^{cl}f^{(0)}=0
\ee
and
\be
{\cal L}^{cl}f^{(2)}+{\cal L}^{pn}f^{(0)}=0.
\ee
The first of these equations implies that
\be
\label{ordenceroliouville}
\frac{\partial f^{(0)}}{\partial t}+{v^i}\frac{\partial f^{(0)}}{\partial x^i}=
\frac{\partial \phi}{\partial x^i}\frac{\partial f^{(0)}}{\partial v^i}.
\ee
Moreover,
\be
{\cal L}^{cl}f^{(2)}=\frac{\partial f^{(2)}}{\partial t}+{v^i}
\frac{\partial f^{(2)}}{\partial x^i}-\frac{\partial \phi}{\partial x^i}
\frac{\partial f^{(2)}}{\partial v^i}\label{B5}
\ee
whereas
\bea
{\cal L}^{pn}f^{(0)}=&&\!\!\!\!\!\!\!\!
\frac{1}{c^{2}}\left(\frac{\mathbf{v}^{2}}{2}-\phi \right)\left(\frac{\partial f^{(0)}}{\partial t}
+v^{i}\frac{\partial f^{(0)}}{\partial x^{i}}\right)\nonumber\\
&&\!\!\!\!\!\!\!\!+\frac{1}{c^{2}}\left[4v^{i} v^{j}\frac{\partial \phi}{\partial x^{j}}-
\left(\frac{3\mathbf{v}^{2}}{2}+3\phi\right)\frac{\partial \phi}{\partial x^{i}}+3v^{i}\frac{\partial \phi}{\partial t}-\frac{\partial \psi}{\partial x^{i}}-
\frac{\partial \xi_{i}}{\partial t}
-v^{j}\left(\frac{\partial \xi_{i}}{\partial x^{j}}-\frac{\partial \xi_{j}}{\partial
  x^{i}}\right)\right]\frac{\partial f^{(0)}}{\partial v^i}.\nonumber\label{B6}
\eea
Replacing (\ref{ordenceroliouville}) in the previous equation we find
\be
{\cal L}^{pn}f^{(0)}=\frac{1}{c^2}\left[{4v^{i} v^{j}}
\frac{\partial \phi}{\partial x^{j}}-({\mathbf{v}^{2}}+4\phi)
\frac{\partial \phi}{\partial x^{i}}+{3v^i}
\frac{\partial \phi}{\partial t}-\frac{\partial \psi}{\partial x^{i}}-
\frac{\partial \xi_{i}}{\partial t}-v^{j}
\left(\frac{\partial \xi_{i}}{\partial x^{j}}-\frac{\partial \xi_{j}}{\partial x^{i}}\right)
\right]\frac{\partial f^{(0)}}{\partial v^i}.\label{B7}
\ee
Here we need the 1PN equations of motion which, in general, are given by \cite{WB}
\be
\frac{dv^i}{dt}=-\frac{\partial \phi}{\partial x^i}+
\frac{1}{c^2}\left[{4v^{i} v^{j}}\frac{\partial \phi}{\partial x^{j}}-
({\mathbf{v}^{2}}+4\phi)\frac{\partial \phi}{\partial x^{i}}+{3v^i}\frac{\partial \phi}{\partial t}-
\frac{\partial \psi}{\partial x^{i}}-\frac{\partial \xi_{i}}{\partial t}-v^{j}
\left(\frac{\partial \xi_{i}}{\partial x^{j}}-
\frac{\partial \xi_{j}}{\partial x^{i}}\right)\right].\label{1PN-comp}
\ee
Introducing (\ref{1PN-comp}) in (\ref{B7}) we obtain
\be
{\cal L}^{pn}f^{(0)}=\left[\frac{dv^i}{dt}+
\frac{\partial \phi}{\partial x^i}\right]\frac{\partial f^{(0)}}{\partial v^i},
\ee
so
\be
({\cal L}^{cl}+{\cal L}^{pn})f^{(0)}=\frac{\partial f^{(0)}}{\partial t}+
{v^i}\frac{\partial f^{(0)}}{\partial x^i}+\frac{dv^i}{dt}\frac{\partial f^{(0)}}{\partial v^i}.
\ee
Also, note that using the 1PN equations of motion in (\ref{B5}), we only have to keep the leading term:
\be
{\cal L}^{cl}f^{(2)}=\frac{\partial f^{(2)}}{\partial t}+
{v^i}\frac{\partial f^{(2)}}{\partial x^i}+\frac{dv^i}{dt}\frac{\partial f^{(2)}}{\partial v^i}.
\ee
Putting all of this together we can finally write
\be
\begin{split}
{\cal L}_v f=&\;{\cal L}^{cl}f^{(0)}+({\cal L}^{cl}f^{(2)}+{\cal L}^{pn}f^{(0)})\\
=&\;\left(\frac{\partial }{\partial t}+{v^i}\frac{\partial}{\partial x^i}+
\frac{dv^i}{dt}\frac{\partial}{\partial v^i}\right)(f^{(0)}+f^{(2)})=0,
\end{split}
\ee
or
\be
\frac{df}{dt}=0.\label{total}
\ee
Now, if we use a set of canonical coordinates
$\mathbf{w}=(\mathbf{q},\mathbf{p})$ (i.e. a set of coordinates
that satisfies the Hamilton equations), we can rewrite (\ref{total}) as
\be
\frac{df}{dt}=\frac{\partial f}{\partial t} +\frac{dq^i}{dt}
\frac{\partial f}{\partial q^i}+\frac{dp^i}{dt}\frac{\partial f}{\partial p^i}
=
\frac{\partial f}{\partial t}+\{f,H\}=0,
\ee
or
\be
\frac{df}{dt}=
\frac{\partial f}{\partial t}+\frac{\partial}{\partial q^i}
\left(f \frac{dq^i}{dt}\right)+\frac{\partial}{\partial p^i}\left(f\frac{dp^i}{dt}\right)=0,
\ee
which means that
\be
\frac{\partial f}{\partial t}+\frac{\partial}{\partial w^m}\left(f \frac{dw^m}{dt}\right)=0,
\ee
where $m=1,...,6$ and $w^m$ is the $m$-th component of $\mathbf{w}$.

\section{Derivation of Equations (\ref{eqs-campo-explic1})-(\ref{eqs-campo-explic2})\label{sec:ap1}}

In order to calculate the r.h.s of equations (\ref{ecs-campo}), we have to write the
required components of the energy-momentum tensor,  $^{0}{T^{00}}$,
$^{2}{T^{00}}$ and $^{2}{T^{ii}}$ (sum over $i=1,2,3$)
explicitly
\begin{eqnarray}
^{0}{T^{00}}
&=&2 k_{n}\int_{0}^{\sqrt{-2\phi}}\left(-\frac{\mathbf{v}^{2}}{2}-\phi\right)^{n-3/2}\mathbf{v}^{2}dv,\label{T000}\\
^{2}{T^{00}}
&=&\frac{6 k_{n}}{c^2} \int_{0}^{\sqrt{-2\phi}}
\left(-\frac{\mathbf{v}^{2}}{2}-\phi\right)^{n-3/2}\left(\mathbf{v}^{2}-2\phi\right)\mathbf{v}^{2}dv\nonumber\\
&&+\frac{2 k_{n}}{c^2}\left(n-3/2\right) \int_{0}^{\sqrt{-2\phi}}
\left(-\frac{\mathbf{v}^{2}}{2}-\phi\right)^{n-5/2}\left(\frac{3\mathbf{v}^{2}\phi}{2}-\frac{3\mathbf{v}^{4}}{8}-\frac{\phi^2}{2}-\psi\right)\mathbf{v}^{2}dv,\nonumber\\
^{2}{T^{ii}}
&=&\frac{2 k_{n}}{c^2}\int_{0}^{\sqrt{-2\phi}}\left(-\frac{\mathbf{v}^{2}}{2}-\phi\right)^{n-3/2}\mathbf{v}^{4}dv.
\end{eqnarray}
The above formulae can be computed easily by making the substitution $\mathbf{v}^{2}=-2\phi \cos^{2}\theta$
and running the integrals over $\theta$,  from $0$ to $\pi/2$. The resulting trigonometric
 integrals can be expressed in terms
of gamma functions by using the relation \cite{abra}
\be
\int_{0}^{\pi/2}(\cos\theta)^{2m}(\sin\theta)^{2k+1} d\theta=
\frac{\Gamma(k+1)\Gamma(m+1/2)}{2\Gamma(m+k+3/2)},
\ee
for $m>-1/2$ and $k>-1$. Then we obtain, for $n>1/2$,
\be
\begin{split}
^{0}{T^{00}}=&\;2k_{n}I(\phi,n-3/2,1),\\
^{2}{T^{00}}+~\!\!^{2}{T^{ii}}=&\;\frac{k_{n}}{c^2}\bigg\{8I(\phi,n-3/2,2)
-12\phi I(\phi,n-3/2,1)+ (n-3/2)
\left[3\phi I(\phi,n-5/2,2)\right.\\
&\quad\quad\;\;\left.-(\phi^{2}+2\psi)I(\phi,n-5/2,1)-\,(3/4)I(\phi,n-5/2,3)\right]\bigg\},
\end{split}
\ee
where we have introduced the notation
\begin{equation}\label{def-I}
    I(\phi,k,m)=\frac{2^{m-1/2}\Gamma(k+1)\Gamma(m+1/2)}{\Gamma(m+k+3/2)}(-\phi)^{k+m+1/2},
\end{equation}
and after some calculations we obtain the expressions (\ref{eqs-campo-explic1}-\ref{eqs-campo-explic2}).

\end{document}